 \documentclass[12pt,preprint]{aastex}

 \usepackage{emulateapj5}

\slugcomment{to appear in the Astrophysical Journal}

\shorttitle{RX~J0513.9$-$6951}
\shortauthors{Hachisu and Kato}

\begin{document}

\title{\objectname{RX~J0513.9$-$6951}: THE FIRST EXAMPLE OF ACCRETION
WIND EVOLUTION, A KEY EVOLUTIONARY PROCESS TO TYPE I\lowercase{a}
SUPERNOVAE}


\author{Izumi Hachisu}
\affil{Department of Earth Science and Astronomy, 
College of Arts and Sciences, University of Tokyo,
Komaba, Meguro-ku, Tokyo 153-8902, Japan} 
\email{hachisu@chianti.c.u-tokyo.ac.jp}

\and

\author{Mariko Kato}
\affil{Department of Astronomy, Keio University, 
Hiyoshi, Kouhoku-ku, Yokohama 223-8521, Japan} 
\email{mariko@educ.cc.keio.ac.jp}




\begin{abstract}
     A new self-sustained model for long-term light curve variations
of \objectname{RX~J0513.9$-$6951} is proposed based on an optically
thick wind model of mass-accreting white dwarfs (WDs).
When the mass accretion rate to a WD exceeds the critical rate of 
$\sim 1 \times 10^{-6} M_\sun$~yr$^{-1}$,
optically thick strong winds begin to blow from the WD so that
a formation of common envelope is avoided.  The WD can accrete and
burn hydrogen-rich matter atop the WD at the critical rate.  The excess
matter transferred to the WD above the critical rate is expelled
by winds.  This is called the accretion wind evolution. 
This ejection process, however, occurs intermittently
because the mass transfer is attenuated by strong winds:
the strong winds collide with the secondary surface and
strip off the very surface layer of the secondary.
The matter stripped-off is lost from the binary system.
Properly formulating this mass stripping effect and the ensuing
decay of mass transfer rate, we are able to reproduce, in a
self-sustained manner, 
the transition between the optical high/X-ray off and
optical low/X-ray on states of \objectname{RX~J0513.9$-$6951}.  
Thus \objectname{RX~J0513.9$-$6951} is the first example 
of the accretion wind evolution, which is a key evolutionary process
in a recently developed evolutionary path to Type Ia supernovae.
\end{abstract}




\section{INTRODUCTION}
     The Large Magellanic Cloud (LMC) transient supersoft X-ray 
source \objectname{RX~J0513.9$-$6951} (hereafter RX~J0513) was 
discovered by ROSAT \citep{sch93} and has been extensively observed
\citep[e.g.,][for recent progress]{rei00}.
Soon after the X-ray discovery, it was optically
identified as a binary with an orbital period of 0.76 days 
\citep{pak93}.  Its remarkable observational features
are summarized as follows: (1) Optical monitoring of RX~J0513 shows
quasi-regular transitions between high ($V \sim 16.6$) and low
($V \sim 17.4$) states \citep{alc96}.  
(2) The duration of the optical high state
is $\sim 60-150$ days while the low states last for $\sim 40$ days
\citep[see, e.g., Fig. 1 of][for recent observational summary]{cow02}.
(3) Copious supersoft X-rays ($\sim 30-40$ eV) were detected only in
the optical low state \citep[e.g.,][]{sch93, rei96, sou96}.
(4) The transitions between X-ray off and on are very rapid,
in about one day or so, while the optical 
transitions from high to low occurs in several days 
\citep[e.g.,][]{rei00}.  The transition mechanism between
high and low states has not been fully elucidated yet, although
a few ideas were proposed so far
\cite[see, e.g.,][for a recent progress]{rei00}.
\par
     Very recently, based on an optically thick wind model of 
mass-accreting white dwarfs (WDs), 
\citet{hac03k} have proposed a new transition mechanism 
between optical high/X-ray off and optical low/X-ray on states.
They first modeled light curves of the 2000 outburst 
of the recurrent nova \objectname{CI Aql} and then showed that 
the size of a disk around the WD expands largely up 
to the companion star or over in a strong wind phase.
Including this effect of disk expansion in the strong wind phase,
they have reproduced a sharp $\sim 1$ mag rise in the optical light
curve of RX~J0513.  This is the optical high/X-ray off state, because 
the optically thick winds of WDs certainly obscure supersoft X-rays. 
On the other hand, it drops sharply by a magnitude 
when the wind stops because the size of the disk returns 
to its original size.  This is the optical low/X-ray on state.  
However, \citet{hac03k} gave only a simple outline of their
mechanism without detailed physical descriptions.  In this paper,
we propose a full description of the mechanism and calculate
light curves for various conditions.  We are able to reproduce 
basic features of the observed light curves by this new mechanism.  
\par
     It has been frequently discussed that supersoft X-ray sources 
(SSSs) are progenitors of Type Ia supernovae (SNe Ia).
One of the promising evolutionary models to SNe Ia has been proposed
by \citet*{hkn96}.
They theoretically found that mass accreting WDs blow strong winds
when the mass accretion to the WD exceeds a critical rate, i.e., 
$\dot M_{\rm acc} > \dot M_{\rm cr} 
\sim 1 \times 10^{-6} M_\sun$~yr$^{-1}$.  
The so-called common envelope evolution does not occur 
\citep[see][for a recent summary]{hac01kb}.  
The WD accretes hydrogen-rich matter and burns it atop the WD
at the critical rate of $\dot M_{\rm cr}$.  
The excess matter transferred above the critical rate is expelled
by winds, i.e., $\dot M_{\rm wind}= \dot M_{\rm acc} - \dot M_{\rm cr}$.
In such a situation, the WD can grow up to 
near the Chandrasekhar mass (and eventually explode as an SN Ia).
This is called the accretion wind evolution 
\citep*{hac02, hac01kb} 
against the common envelope evolution \citep{ibe84, web84}.  
Thus, the accretion wind evolution is 
a key evolutionary process to produce Type Ia supernovae.
\par
     Binary systems in the accretion wind evolution are predicted
to exist but no such systems have been identified yet.
Here, we have reached the conclusion that 
\objectname{RX~J0513.9$-$6951} is the first example
just in the accretion wind evolution, i.e., in the way to
an SN Ia, the reason of which will be discussed below.
A brief summary follows: The wind ejection process occurs not 
continuously but intermittently because the mass transfer
is attenuated by the strong wind itself.
\par
     In \S 2, an outline of our basic model of RX~J0513 is briefly
introduced and summarized.  
Our new mechanism of transition between the optical 
high and low states, in which the mass transfer rate is attenuated by
winds, is formulated in \S 3.  A limit cycle model is given in \S 4 
for RX~J0513 long-term light curve variations.
Discussion and conclusions follow in \S5 and \S6, respectively.

\placefigure{rxj0513_fig}

\section{BRIEF DESCRIPTION OF A LIMIT CYCLE MODEL}
     Long-term variations in the optical light curves of 
\objectname{RX~J0513.9$-$6951} are considered as a limit cycle 
\citep[e.g.,][]{rei00}.
Before going to detailed descriptions, we here give a rough 
sketch on our new mechanism.  We adopt a binary system consisting of 
a WD, a disk around the WD, and a lobe-filling main-sequence (MS)
companion, as illustrated in Figure \ref{rxj0513_fig}.  
A circular orbit is assumed. 
\par
     An essential thing is that WDs blow an optically thick, strong 
wind when the mass accretion rate to the WD exceeds the critical 
rate, i.e., $\dot M_{\rm acc} > \dot M_{\rm cr} \sim 1 \times 10^{-6} 
M_\sun$~yr$^{-1}$.  If the MS companion is $\sim 2-3 ~M_\sun$, 
the mass transfer rate from the MS, $\dot M_{\rm MS}$, 
is as large as a few times $10^{-6} M_\sun$~yr$^{-1}$ or more
\citep[e.g.,][]{lih97}.
We distinguish $\dot M_{\rm MS}$ from $\dot M_{\rm acc}$
because $\dot M_{\rm MS} \ne \dot M_{\rm acc}$ in some cases 
mentioned below.  
\par
     Time-evolution of our limit cycle model is schematically
summarized in Figure \ref{vmagfit_rxj0513_self_one} 
\citep[see also][]{hac03k}.
A rapid mass accretion to the WD begins at A in Figure
\ref{vmagfit_rxj0513_self_one}.
The mass of the WD envelope rapidly increases and
the WD envelope expands to blow an optically thick, strong wind.
The strong wind causes a quick rise of $V$-mag at B, 
because the disk expands to a few to several times 
the previous size, that is, the disk expands from configuration (c)
to configuration (a) in Figure \ref{rxj0513_fig}.
Because the disk is the main source in the optical light due to
its irradiation effect, this expansion of the disk area causes 
a magnitude rise in the optical light.
The massive winds having the rate of 
$\dot M_{\rm wind} \sim 1 \times 10^{-6} M_\sun$~yr$^{-1}$
certainly absorb supersoft X-rays.  Thus, this phase is 
the optical high and X-ray off state. As the WD photosphere
expands further during the rapid mass accretion phase,
optical brightness increases further from B to C 
even if the disk shape does not change in the wind phase. 
\par
     The strong wind hits the MS companion and then strips-off 
its very surface layer.  The gas stripped-off is lost from the
binary.  Therefore, the rapid mass accretion is 
suppressed by the strong wind itself.  The mass-stripping rate is 
roughly proportional to the wind mass loss rate.
As the wind mass loss rate increases,
the mass-stripping rate eventually overcomes the mass transfer rate.
This happens at C'.  However, the accretion disk still continues
to supply mass to the WD at least during the viscous timescale
of the accretion disk,
$t_{\rm vis}$.  Then, mass accretion to the WD stops at C
about $t_{\rm vis}$ after the mass transfer from the MS stops.  
\par
     The mass of the WD envelope is now gradually decreasing 
due mainly to wind mass loss.  The photosphere of the WD is 
also gradually shrinking.  This makes a gradual decay of 
the optical light curve even if the disk shape is the same.
The wind eventually stops at D.
We expect copious supersoft X-rays after the wind stops.
The disk shrinks to a normal size at E, that is,
from configuration (a) to configuration (b) in Figure 
\ref{rxj0513_fig}.   This causes a sharp $\sim 1$ mag drop 
in the optical light.  A rapid mass transfer from the MS
to the disk resumes at E, and then the disk edge
flares up at F, that is, configuration (b) to configuration (c)
in Figure \ref{rxj0513_fig}.  It takes several dynamical timescales
to change from E to F.  
This phase corresponds to the optical low and X-ray on state.
\par
     Because mass accretes to the WD via an accretion disk, 
this rapid mass accretion flow reaches the WD surface 
at G(=A) about $t_{\rm vis}$ after the rapid mass transfer
from the MS resumes.  Thus, the cycle is repeated.

\placefigure{vmagfit_rxj0513_self_one}

\section{THE BINARY MODEL}

\subsection{Basic Parameters of the Binary}
     A relatively massive 
white dwarf is suggested because of a rapid contraction timescale 
of the WD photosphere during the transition from optical high to low
states \citep[e.g.,][]{sou96}.  We have adopted $M_{\rm WD}= 1.3~M_\sun$
after \citet{hac03k} but later examined other cases of the WD mass,
i.e., $M_{\rm WD}= 1.35~M_\sun$,  $1.2~M_\sun$, and  $1.1~M_\sun$.
Since the LMC metallicity is reported to be about a third of
the solar metallicity, we assume the metallicity of $Z=0.004$ and
hydrogen content of $X=0.7$ for the envelope of the WD.
The mass ratio is assumed to be $q=M_{\rm MS} / M_{\rm WD} =2$
because mass transfer occurs on a thermal timescale and a rate 
of $\dot M_{\rm MS} \gtrsim 10^{-6} M_\sun$~yr$^{-1}$
is suggested by \citet{sou96}.
Then, the companion mass is $M_{\rm MS} = 2.6 M_\sun$
for $M_{\rm WD}= 1.3~M_\sun$.
The non-irradiated surface temperature of
the lobe-filling $2.6 M_\sun$ MS companion is 
13,000 K for the metallicity of $Z=0.004$ \citep[e.g.,][]{fag94}.
\par
     For the orbital period of $P_{\rm orb} = 0.763$ days 
(see below), the separation is $a= 5.53 ~R_\sun$, the radii of the
effective Roche lobes are $R_1^* = 1.77 ~R_\sun$ for the primary (WD)
and $R_2 = R_2^* = 2.43 ~R_\sun$ for the secondary (MS) component.
The radius of a $2.6 ~M_\sun$ zero-age main-sequence
(ZAMS) star is $R_{\rm ZAMS}= 1.5 ~R_\sun$ for $Z=0.004$, 
so that the secondary has already evolved to expand but still
remains at the main-sequence (before exhaustion of central
hydrogen).

\par
The ephemeris of RX~J0513 has been revised by \citet{cow02} as 
\begin{equation}
t ({\rm JD}) = 2,448,858.099 + 0.7629434 \times E
\label{RXJ0513_ephemeris}
\end{equation}
at minima of the optical orbital modulations.  We use this revised
ephemeris with the MS companion being in front of the WD at minima
\citep[see also Fig. 5 of][]{hut02}.
The inclination angle is not known but suggested to be
low.  Here we adopt $i= 10\arcdeg$.  Calculated light curves are not
so different among $i=0\arcdeg - 35\arcdeg$ as shown later.  

\subsection{Time Evolution of the WD Envelope}
     Time evolution of the photospheric radius and temperature 
of the WD are calculated from Kato \& Hachisu's (1994) 
optically thick wind solutions after \citet{hac01kb}, i.e.,
\begin{equation}
{{d } \over {d t}} \Delta M_{\rm env} = \dot M_{\rm acc} 
- \dot M_{\rm nuc} - \dot M_{\rm wind},
\label{on_white_dwarf_surface}
\end{equation}
where $\Delta M_{\rm env}$ is the hydrogen-rich envelope mass of the WD,
$\dot M_{\rm acc}$ the mass accretion rate to the WD,
$\dot M_{\rm nuc}$ the decreasing rate by nuclear burning, and
$\dot M_{\rm wind}$ the wind mass loss rate of the WD.
The nuclear burning rate $\dot M_{\rm nuc}$, wind mass loss rate
$\dot M_{\rm wind}$, photospheric radius $R_{\rm ph}$, 
photospheric temperature $T_{\rm ph}$, 
and luminosity $L_{\rm ph}$ as well as the photospheric wind velocity
$v_{\rm ph}$, of the WD
are all given as a function of the envelope mass ($\Delta M_{\rm env}$)
in the wind solutions \citep{kat94h, hac01kb}, i.e.,
\begin{equation}
\dot M_{\rm nuc} = f_1(\Delta M_{\rm env}, X, Z, M_{\rm WD}),
\end{equation}
\begin{equation}
\dot M_{\rm wind} = f_2(\Delta M_{\rm env}, X, Z, M_{\rm WD}),
\end{equation}
\begin{equation}
R_{\rm ph} = f_3(\Delta M_{\rm env}, X, Z, M_{\rm WD}),
\end{equation}
\begin{equation}
T_{\rm ph} = f_4(\Delta M_{\rm env}, X, Z, M_{\rm WD}),
\end{equation}
\begin{equation}
L_{\rm ph} = f_5(\Delta M_{\rm env}, X, Z, M_{\rm WD}),
\end{equation}
\begin{equation}
v_{\rm ph} = f_6(\Delta M_{\rm env}, X, Z, M_{\rm WD}).
\end{equation}
Given the mass accretion rate $\dot M_{\rm acc}$ and the initial
envelope mass, we are able to
follow the time evolutions of the envelope mass and then 
the radius and temperature (and also the luminosity) 
of the WD by integrating equation (\ref{on_white_dwarf_surface}). 

\subsection{Optically Thick Strong Winds} 
     Mass transfer in RX~J0513 probably proceeds in a Kelvin-Helmholtz
timescale of the MS companion \citep[e.g.,][]{heu92},
so that it is as large as 
$\dot M_{\rm MS} \sim 1 \times 10^{-6} M_\sun$~yr$^{-1}$ or more.
When the mass transfer rate, $\dot M_{\rm MS}$,  
exceeds the critical rate, i.e.,
\begin{equation}
\dot M_{\rm MS} > \dot M_{\rm cr} \approx 0.75 \times 10^{-6} 
\left( {{M_{\rm WD}} \over 
{M_\sun}} - 0.4 \right)~M_\sun {\rm ~yr}^{-1},
\label{critical_mass_accretion}
\end{equation}
the mass-accreting WD expands to blow optically thick winds
\citep[e.g.,][]{hkn96} and the mass loss rate approximately given by
\begin{equation}
\dot M_{\rm wind} \approx \dot M_{\rm MS} - \dot M_{\rm cr},
\end{equation} 
when all the transferred matter accretes to the WD, i.e.,
$\dot M_{\rm acc} = \dot M_{\rm MS}$.
Equation (\ref{critical_mass_accretion}) is the same form as that for 
the solar metallicity ($Z=0.02$), but valid even for a lower 
metallicity of $Z=0.004$ \citep*[e.g.,][]{hac01kb}.

\placefigure{binary_flow}

\subsection{Mechanism of Intermittent Mass Accretion}
     When strong winds collide with the surface of 
the companion, the surface is shock-heated and ablated in the wind. 
We estimate the shock-heating by assuming that the velocity component
normal to the surface is dissipated by shock and its kinetic energy
is converted into the thermal energy of the surface layer.
The very surface layer of the envelope expands to be easily ablated 
in the wind.  We regard that gas is ablated and lost from L3 point
(L3 is the outer Lagrange point near the MS companion) 
when the gas gets the same amount
of thermal energy as the difference of the Roche potentials between
the MS surface and L3 point. 
Then the mass stripping rate is given by
\begin{equation}
{{G M} \over {a}} \left( \phi_{\rm L3} - \phi_{\rm MS} \right) \cdot 
\dot M_{\rm strip} 
=  {1 \over 2} v^2 \dot M_{\rm wind} \cdot \eta_{\rm eff} \cdot g(q),
\label{strip_off_origin}
\end{equation}
where $M=M_{\rm WD}+ M_{\rm MS}$, $a$ is the separation of the binary,
$\phi_{\rm MS}$ the Roche potential at the MS surface, 
$\phi_{\rm L3}$ means the Roche potential at L3 point
near the MS companion, both of which are normalized by $GM/a$,
$\eta_{\rm eff}$ the efficiency of conversion from kinetic energy 
to thermal energy by shock, 
$g(q)$ is the geometrical factor of the MS surface 
hit by the wind including the dissipation effect (normal 
component of the wind velocity), and $g(q)$ is only 
a function of the mass ratio $q=M_{\rm MS}/M_{\rm WD}$ 
\citep*[see][for more details]{hkn99}.  Here we modified 
equation (21) of \citet{hkn99} to include the effect of Roche lobe
overflow from L3 point.  Then the stripping rate is estimated as
\begin{equation}
\dot M_{\rm strip} = c_1 \dot M_{\rm wind},
\label{mass_stripping_rate}
\end{equation}
where
\begin{equation}
c_1 \equiv {{\eta_{\rm eff} \cdot g(q)} \over
{\phi_{\rm L3} - \phi_{\rm MS}}}
\left({{v^2 a} \over {2 G M}} \right).
\end{equation}
The efficiency of conversion is assumed to be $\eta_{\rm eff}=1$. 
Then, we have
\begin{equation}
c_1 \approx
0.1 \left( {v \over {400~{\rm km~s}^{-1}}} \right)^2 \approx  10,
\label{mass_stripping_rate_real}
\end{equation}
for the wind velocity of $v_{\rm wind}= 3800$~km~s$^{-1}$
\citep[e.g.,][]{cra96, hut02, sou96},  
$a = 5.5 R_\sun$, $M= 3.9 M_\sun$,
$\phi_{\rm L3} - \phi_{\rm MS} = 0.3$,
and $g(q)=0.025$ for the mass ratio 
$q= M_{\rm MS}/M_{\rm WD} = 2$ \citep{hkn99}.
\par
     In such a wind phase, the net mass accretion rate to the WD, 
$\dot M_{\rm acc}$, is modified as 
\begin{equation}
\dot M_{\rm acc} = \left\{
        \begin{array}{@{\,}ll}
          0~(\mbox{~or~}\epsilon), 
& \quad \mbox{~for~} \dot M_{\rm MS} \le \dot M_{\rm strip} \cr
          \dot M_{\rm MS} - \dot M_{\rm strip},
& \quad \mbox{~for~} \dot M_{\rm MS} > \dot M_{\rm strip}
        \end{array}
      \right.
\label{accretion_rate}
\end{equation}
as illustrated in Figure \ref{binary_flow}.  
Here we adopt a small value of 
$\epsilon= 1 \times 10^{-7} M_\sun$~yr$^{-1}$  
when $\dot M_{\rm MS} < \dot M_{\rm strip}$ because the mass
accretion to the WD does not stop abruptly but probably continues
at least for a draining time of the accretion disk.  We do not know
the exact draining time of the accretion disk after the mass 
transfer from the MS stops.  Alternatively, we just assume a small rate
of the mass accretion $\epsilon$ to mimic the draining of the disk
during that $\dot M_{\rm MS} < \dot M_{\rm strip}$.
\par
     To know when the rapid mass accretion resumes,
we monitor the upper level of flow in Figure \ref{binary_flow} as
\begin{equation}
{{d } \over {d t}} M_{\rm flow}= \dot M_{\rm MS} - \dot M_{\rm strip}
- \dot M_{\rm acc},
\label{on_secondary_surface}
\end{equation}
with the initial value of $M_{\rm flow}=0$.  
During that $\dot M_{\rm strip} > \dot M_{\rm MS}$, 
$M_{\rm flow}$ decreases to a large negative value
as illustrated in Figure \ref{binary_flow}b.
Then the mass transfer from the MS stops or its rate drops 
to $\epsilon$.  Once the mass accretion rate drops below 
the critical rate, i.e., $\dot M_{\rm acc} < \dot M_{\rm cr}$,
the wind gradually weakens and eventually stops.
The stripping effect vanishes ($\dot M_{\rm strip}= 0$) 
and then the level of flow begins to rise.
We start again the rapid mass accretion when the level reaches
$M_{\rm flow}=0$ as illustrated in Figure \ref{binary_flow}a.
\par
     It should be noticed that equation (\ref{on_white_dwarf_surface})
is on the WD but equations 
(\ref{strip_off_origin})$-$(\ref{on_secondary_surface}) 
are on the MS.
Therefore, the mass accretion rate, $\dot M_{\rm acc}$, may not be
the same on each side at the specific time, $t$,
because it takes a viscous timescale of the accretion disk for gas
to reach the WD surface from the MS surface.  Here we adopt 
a viscous timescale of
\begin{equation}
t_{\rm vis} = {{R_{\rm disk}^2} \over {\nu}} = {{R_{\rm disk}} \over
\alpha_{\rm vis} } {{R_{\rm disk}} \over {c_{\rm s} H}}
\sim 40 \left({{\alpha_{\rm vis}} \over {0.1} } \right)^{-1} \mbox{~days},
\label{viscous_timescale}
\end{equation}
where $\nu$ is the viscosity, $\alpha_{\rm vis}$ the $\alpha$ parameter
of \citet{sha73}, $c_{\rm s}$ the sound speed, and $H$ the vertical height
of the accretion disk.  We adopt $R_{\rm disk}= 1.4~R_\sun$ 
and $H/R_{\rm disk} \sim  0.1$ for
$T_{\rm disk} \sim 30000-50000$~K and $c_{\rm s} \sim 30$km~s$^{-1}$.
Then, we have
\begin{equation}
\left. \dot M_{\rm acc}(t) \right|_{\rm WD} = \left. 
\dot M_{\rm acc}(t - t_{\rm vis})
\right|_{\rm MS}.
\end{equation}

\subsection{Irradiation Effects of Disk}
\begin{figure}
\plotone{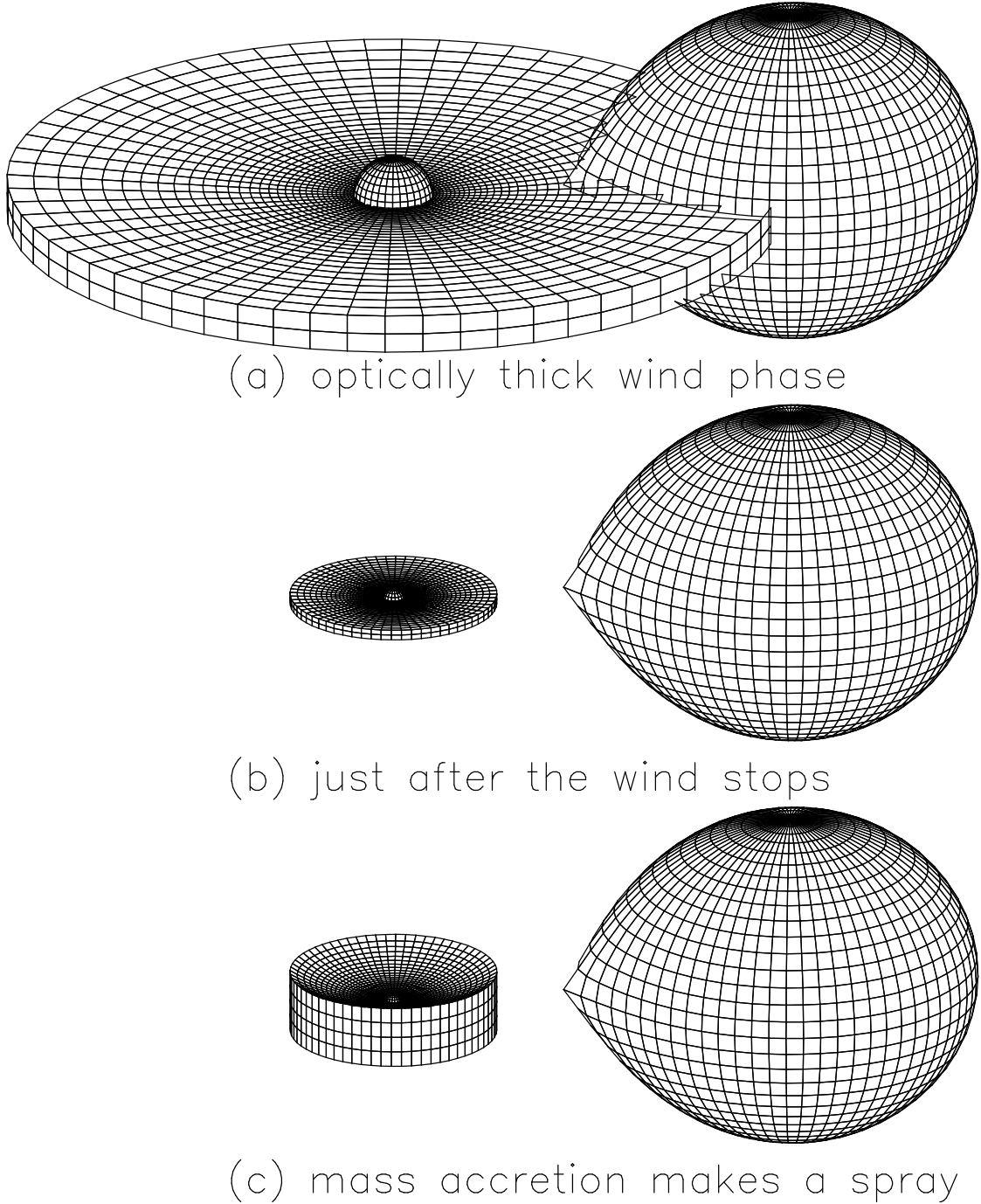}
\caption{
Configurations of our modeled RX~J0513.9$-$6951 are illustrated: (a)
in the massive wind phase (X-ray off), (b) just after the wind stops
(X-ray on), and (c) during a rapid mass accretion phase soon after
the wind stops (X-ray on).  The cool component ({\it right}) is
a slightly evolved main-sequence (MS) companion ($2.6 M_\odot$)
filling its inner critical Roche lobe.  
The north and south polar areas of the cool component are 
irradiated by the hot component ($1.3~M_\odot$ white dwarf, {\it left}).
The separation is $a= 5.53 R_\odot$; 
the effective radii of the inner critical Roche lobes are
$R_1^*= 1.77 R_\odot$, and $R_2^*= R_2= 2.43 R_\odot$, 
for the primary WD and the secondary MS companion, respectively.
(a) The surface of the accretion disk is blown in the wind and
its optically thick outer edge extends up to the companion star or over.
The large velocity difference between the wind and the disk
surface certainly drives the Kelvin-Helmholtz instability
at the interface, so that the very surface layer of the disk is
dragged away as an almost free stream going outward.
This surface free stream is initially optically thick but becomes 
optically thin outside because of geometrical dilution effect.
We regard the transition from optically thick to thin as the edge of
the disk.  
(b) The disk shrinks to a normal size ($0.7 - 0.8$ times the Roche lobe 
radius) in several dynamical timescales, i.e., several orbital periods
after the wind stops.  
(c) A rapid mass accretion makes a spray around the disk edge.
\label{rxj0513_fig}}
\end{figure}

\begin{figure}
\plotone{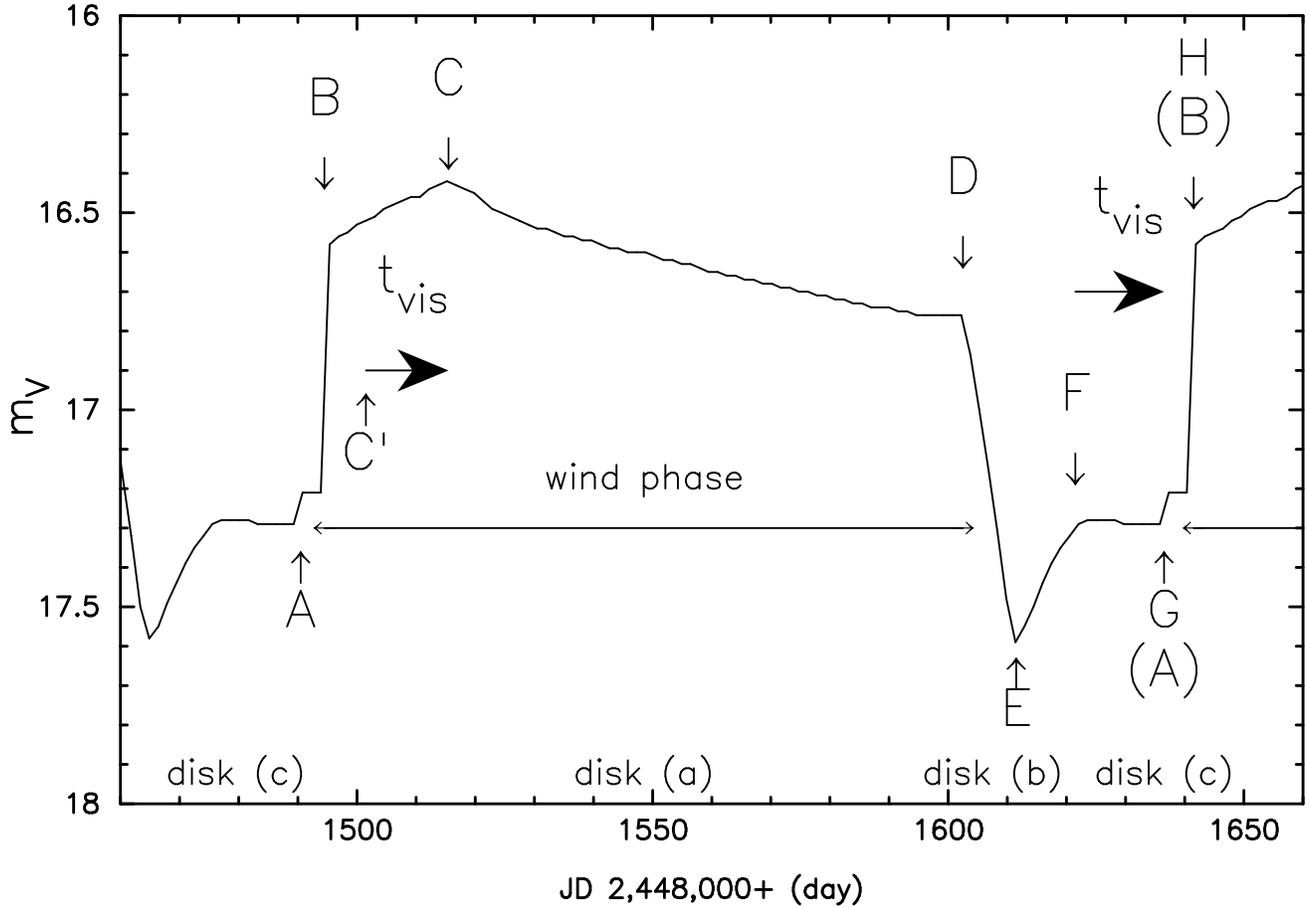}
\caption{
A modeled light curve for a limit cycle is plotted.
A rapid mass accretion to the WD begins at A.
The envelope of the WD expands to blow an optically thick
wind at B.  The strong wind causes a quick rise of 
the optical light curve, because the disk expands
to a few to several times the previous size (see disk configuration
(a) in Fig. \ref{rxj0513_fig}).
The massive wind certainly absorb supersoft X-rays. 
Thus, this is the optical high/X-ray off state.
Because the wind hits and then strips-off the very surface layer 
of the MS companion, mass transfer from the MS is heavily 
suppressed and stops at C'.  However, the mass accretion to the WD
still continues during a viscous timescale of the accretion disk,
i.e., $t_{\rm vis}$.  After a large part of the mass in the accretion
disk is drained, the mass accretion to the WD virtually stops at C.
The mass of the WD envelope is now gradually decreasing due mainly 
to wind mass loss.  Then the WD photosphere is also gradually shrinking.
This makes a slow decline in the optical light curve.
The wind eventually stops at D.  
The disk shrinks to a normal size at E, causing a sharp
drop of the optical light curve (see disk configuration (b) in
Fig. \ref{rxj0513_fig}).
The mass transfer from the MS resumes at E, and then
the edge of the disk flares up at F 
(see disk configuration (c) in Fig. \ref{rxj0513_fig}).
It takes several dynamical timescales to change from E to F.
We expect copious supersoft X-rays after the wind stops.
This is the optical low/X-ray on state.
A rapid mass accretion to the WD resumes at G(=A), because
it takes a viscous timescale, i.e., $t_{\rm vis}$, that the matter
reaches the WD surface via an accretion disk.  This cycle is repeated.
\label{vmagfit_rxj0513_self_one}}
\end{figure}

\begin{figure}
\plotone{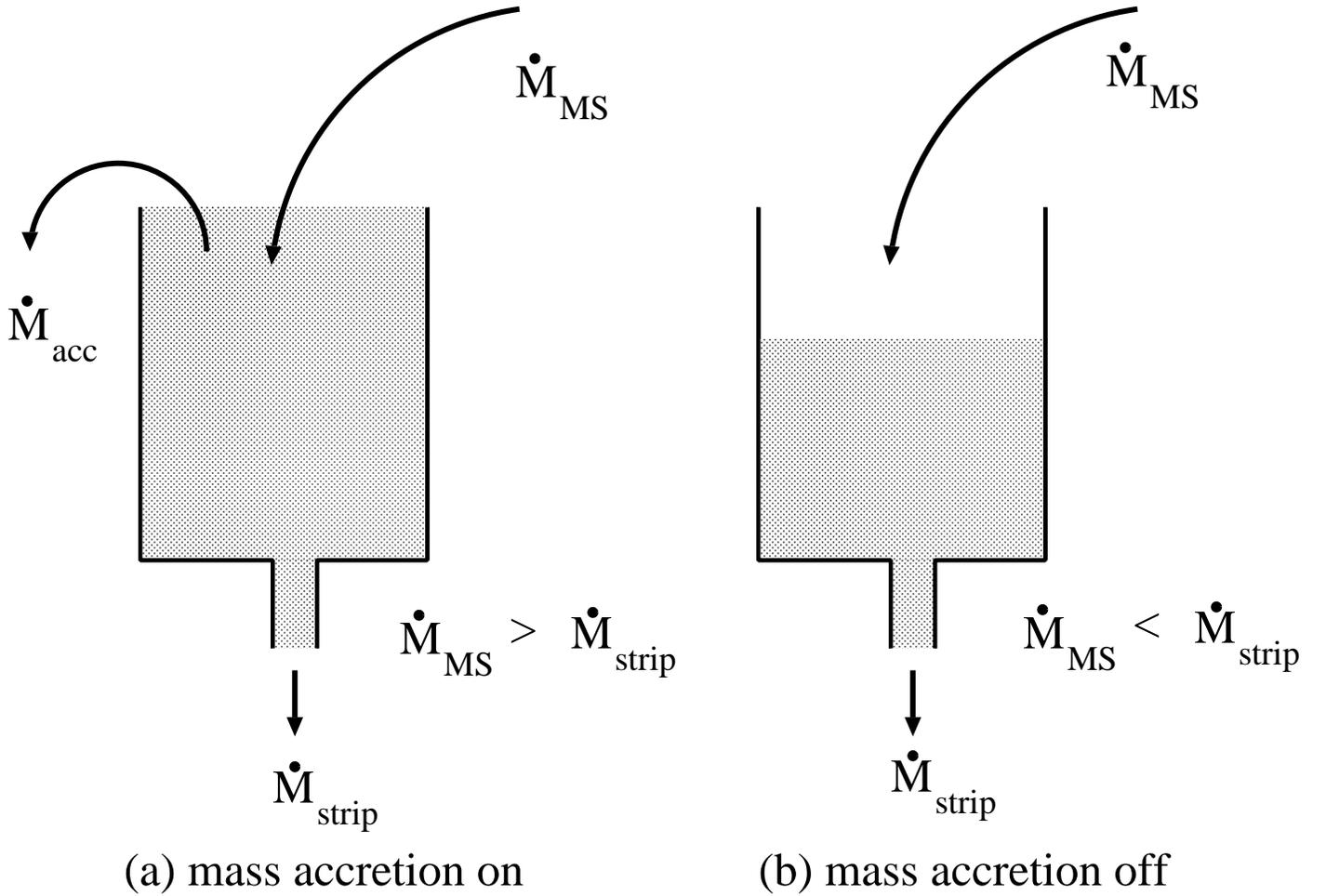}
\caption{
Schematic illustrations for the flow from the secondary main-sequence
companion: (a) mass transfer is going on when the stripping effect
by the wind is smaller than the original mass transfer rate, i.e.,
$\dot M_{\rm strip} < \dot M_{\rm MS}$.  (b) Mass transfer stops
if the stripping effect by the wind is larger than the original
mass transfer rate, i.e., $\dot M_{\rm strip} \ge \dot M_{\rm MS}$.
We are able to know when mass transfer resumes again by checking
whether or not $M_{\rm flow} \ge 0$ by integrating 
equation (\ref{on_secondary_surface}), where $M_{\rm flow}$ indicates
the level of flow ($M_{\rm flow}=0$ at full).
\label{binary_flow}}
\end{figure}

     The irradiation effects both of the disk and of the MS companion
play an essential role in the light curve modeling. 
For the basic structure of the accretion disk,
we assume an axi-symmetric structure with the size and thickness of
\begin{equation}
R_{\rm disk} = \alpha R_1^*,
\label{accretion-disk-size}
\end{equation}
and
\begin{equation}
h = \beta R_{\rm disk} \left({{\varpi} 
\over {R_{\rm disk}}} \right)^\nu,
\label{flaring-up-disk}
\end{equation}
where $R_{\rm disk}$ is the outer edge of the accretion disk,
$R_1^*$ is the effective radius of the inner critical Roche lobe 
for the WD component,
$h$ is the height of the surface from the equatorial plane, and
$\varpi$ is the distance on the equatorial plane 
from the center of the WD.  
Such binary models are illustrated in Figure \ref{rxj0513_fig}.
We assume that the photosphere of the WD, the MS companion, 
and the accretion disk emit photons as a blackbody at a local
temperature which varies with position.
\par
     We adopt $\nu=1$ during the strong wind phases, i.e.,
configuration (a) in Figure \ref{rxj0513_fig} and
just after the wind stops (but before a rapid mass transfer
from the MS resumes), i.e., configuration (b) in Figure 
\ref{rxj0513_fig}, but $\nu=2$ in the no wind case, i.e.,
configuration (c) in Figure \ref{rxj0513_fig}.
This $\varpi$-square law mimics the effect of 
elevated rim of the accretion disk by spray as discussed
by \citet*{sch97} for luminous supersoft X-ray sources.
On the other hand, we simply assume $\nu=1$ in the wind phase,
because such an elevated rim is blown in the wind during 
a strong wind phase.  The efficiency of irradiation is 
assumed to be $0.5$ both for the disk and for the companion
\citep*[see also][]{sch97, mey97}.
\par
     The size and thickness of disk, $\alpha$ and $\beta$, are 
also the important parameters
that determine the brightness of disk by irradiation effects.
\citet{hac03k} have determined $\alpha$ and $\beta$ from the light
curve fitting of \objectname{CI Aql} in outburst.
The disk expands up to the companion
star as shown in Figure \ref{rxj0513_fig}a, i.e., $\alpha \sim 2-3$,
because the disk surface is blown in the wind.
It shrinks to a normal size of $\alpha= 0.7-0.8$
after the wind stops.
These two parameters of $(\alpha, \beta)$ 
are determined as $(\alpha, \beta) = (3.0, 0.05)
\rightarrow (0.8, 0.05) \rightarrow (0.8, 0.3)$, during
the wind phase, just after the wind stops, and then the rapid mass
transfer resumes, respectively \citep[see][for more details]{hac03k}.
The shrink from a large, extended size to a normal size is important
in the modeling of a sharp $\sim 1$ mag drop in the optical light
curve.  Thus, a large expansion of the disk in the strong wind phase
has been first confirmed in the 2000 outburst of the recurrent
nova \objectname{CI Aql}.  
\par
     We find essentially the same sets of $(\alpha, \beta)$ 
as in \citet{hac03k} is valid for RX~J0513.
We may understand such a large expansion
of the disk beyond the Roche lobe as follows:  
Observationally very fast
winds have been reported, for example, $3800$ km~s$^{-1}$ 
in RX~J0513 \citep[see, e.g.,][]{cra96, hut02, sou96}.  
The wind mass loss rate
is calculated to be $10^{-6} - 10^{-4} M_\sun$~yr$^{-1}$ in such 
a strong wind phase \citep[e.g.,][]{hac01kb}.  
There is a large velocity difference between the wind and the disk
surface.  It certainly drives the Kelvin-Helmholtz instability
at the interface, so that the very surface layer of the disk is
dragged away outward with the velocity at least several hundred 
km~s$^{-1}$, i.e., almost a free stream going outward.
This surface free stream is initially optically thick
at the original disk region but becomes optically thin somewhere
outside because of geometrical dilution effect.
Only the optically thick region of disk stream can contribute 
to the optical light by irradiation, so that
we regard the radius of transition region between optically thick
and thin as the edge of the disk.  
In this sense, the disk edge as shown in Figure \ref{rxj0513_fig}a
is not the boundary of matter but just
a transition from optically thick to thin of the disk surface flow.
\par
     It should be noted that a high density part of the disk is hardly
changed by this relatively tenuous surface free stream and 
resides in within its Roche lobe because the internal density
of the disk is much denser than that of the WD wind.  In the case of
Figure \ref{vmagfit_rxj0513_self}, for example, 
the wind mass loss rate is about $1 \times 10^{-6} M_\sun$~yr$^{-1}$
and its velocity is $\sim 4000$ km~s$^{-1}$, so that the density of 
the wind is estimated from the continuity ($\dot M_{\rm wind} = 
4 \pi r^2 \rho v$) to be about $10^{-11}$g~cm$^3$ at the distance 
of $1~R_\sun$ from the center of the WD.
On the other hand, the density of the standard accretion disk
is about $10^{-1}$g~cm$^3$ at the same radius. 
Here, we assume the standard accretion disk model \citep{sha73}
with the mass accretion rate of 
$\sim 5 \times 10^{-6} M_\sun$~yr$^{-1}$.
\par
     It should be also addressed that optically thick winds are 
accelerated deep inside the photosphere ($\tau \gg 1$) 
but the wind itself becomes optically thin ($\tau \ll 1$) 
far above the WD photosphere as the definition of photosphere.
Therefore, we do not expect any large optical contribution from
the wind itself far outside the WD photosphere
\citep[see, e.g.,][for an estimation of small 
wind contributions to the optical light]{mey97}.
\par
     The numerical methods for obtaining light curves, 
including the above irradiation effects,
have been described in more detail in \citet{hac01kb}. 
\placefigure{vmagfit_rxj0513_self} 

\section{LIMIT CYCLE MODEL OF LONG-TERM VARIATIONS}

\subsection{A template light curve}
\begin{figure}
\plotone{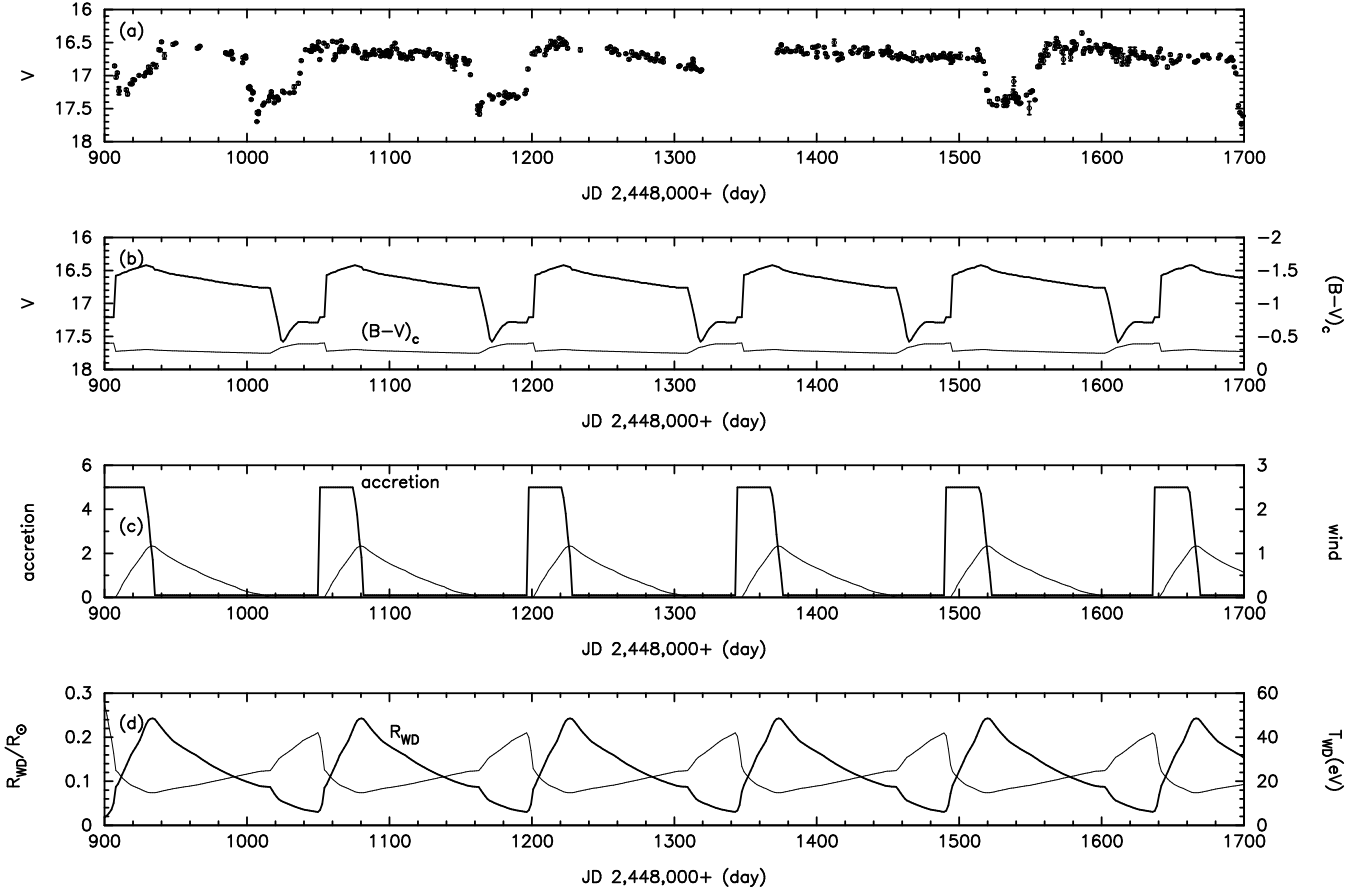}
\caption{
Numerical results of our RX~J0513.9$-$6951 models are shown
against time (JD 2,448,000+) together with observational 
light curve:
(a) Observational $V$-magnitudes \citep[taken from][]{alc96}.
(b) Calculated $V$-magnitudes ({\it thick solid line}) together 
with $B-V$ color ({\it thin solid line}).
(c) Mass accretion rate to the WD ($\dot M_{\rm acc}$, 
{\it thick solid}) and wind mass loss rate from the WD 
($\dot M_{\rm wind}$, {\it thin solid}), both
in units of $10^{-6} M_\sun$yr$^{-1}$.
(d) Photospheric radius of the WD envelope in units of $R_\sun$
({\it thick solid}) and surface temperature of the WD envelope 
in units of eV ({\it thin solid}).
The model parameters are summarized in Table \ref{high_low_states},
i.e., $M_{\rm WD}= 1.3~M_\sun$, $\dot M_{\rm MS}= 5.0 \times
10^{-6}M_\sun$~yr$^{-1}$, $c_1=10.0$, and $t_{\rm vis}= 20.5$ days.
Direct fitting to the brightness of the observational light curves
indicates an apparent distance modulus of $(m-M)_V = 18.7$.
\label{vmagfit_rxj0513_self}}
\end{figure}

Assuming $\dot M_{\rm MS} = 5 \times 10^{-6} M_\sun$~yr$^{-1}$,
$c_1 = 10.0$, and $t_{\rm vis}= 20.5$ days, 
we have calculated time evolution 
of our binary system with the initial envelope mass of
$\Delta M_{\rm env}= 5 \times 10^{-7} M_\sun$.
The obtained $V$-light curve and color (second panel), 
accretion rate $(\dot M_{\rm acc}$ on the WD) 
and wind mass loss rate in units of $10^{-6}M_\sun$~yr$^{-1}$ (third
panel), radius ($R_\sun$) and temperature (eV) of the WD photosphere
(bottom panel) together with the MACHO observation (top panel) 
are shown in Figure \ref{vmagfit_rxj0513_self}.
Direct fitting to the observational light curves
gives us an apparent distance modulus of $(m-M)_V = 18.7$.
If we use the color excess of $E(B-V)= 0.13$ \citep{gan98},
the distance to RX~J0513 is $(m-M)_0 = 18.3$, being consistent
with the relatively short distance to the LMC.  In this paper, we
use $(m-M)_V = 18.7$.
\par
     At the beginning of time-evolution calculations,
the mass of the WD envelope is small so that the WD
does not blow a wind yet.  Its envelope mass increases because of 
$\dot M_{\rm MS} = \dot M_{\rm acc} > \dot M_{\rm nuc}$
($\dot M_{\rm wind}= 0$).  When the envelope mass reaches
$\Delta M_{\rm env} = 6.5 \times 10^{-7} M_\sun$, the WD photosphere
expands to $R_{\rm ph}= 0.087 ~R_\sun$ and begins to blow a wind.
The disk size expands from $(\alpha, \beta) = (0.8, 0.2)$ 
to $(3.0, 0.05)$ in dynamical timescales.  Here we adopt a few days 
as the transition time, because the strong winds easily blow 
the very surface layer of the disk in the wind.
The wind mass loss rate is about $\dot M_{\rm wind}= 1 \times 10^{-8}
M_\sun$~yr$^{-1}$ only one day after the wind starts, which is 
large enough to completely obscure soft X-rays 
\citep[see, e.g., discussion of][]{sou96}.
The occurrence of optically thick winds can naturally explain 
the very rapid emergence and decay ($\lesssim 1-2$ days) 
of supersoft X-rays accompanied by the transition
between the optical high and low states \citep{rei00}.
We discuss again in more detail this transition timescale of 
soft X-rays bellow in \S5.
\par
     The envelope mass continues to increase and the wind mass loss
rate also grows.  The WD photosphere also continues to expand and 
this causes a gradual increase in the optical light curve.
The mass transfer from the MS is suppressed
when the wind mass loss rate reaches 
$\dot M_{\rm wind} = 0.5 \times 10^{-6} M_\sun$~yr$^{-1}$, because
$\dot M_{\rm strip} = 10 \dot M_{\rm wind} = \dot M_{\rm MS} =
5.0 \times 10^{-6} M_\sun$~yr$^{-1}$.
However, the mass accretion to the WD still continues at a high rate
and the wind mass loss rate reaches 
$\dot M_{\rm wind} = 1.0 \times 10^{-6} M_\sun$~yr$^{-1}$ 
as shown in Figure \ref{vmagfit_rxj0513_self}, 
because the mass in the accretion disk
is drained to the WD in another 20.5 days of $t_{\rm vis}$ time delay.
The mass accretion to the WD eventually stops and the mass of 
the WD envelope begins to decrease due to wind mass loss and nuclear
burning.
The WD photosphere gradually shrinks from $R_{\rm WD, ph}= 0.24
~R_\sun$ to $0.087 ~R_\sun$ during the wind phase.  This causes
the gradual decrease in the optical light curve until the end of 
high state.  When the mass of the WD envelope decreases to 
$\Delta M_{\rm env} = 6.5 \times 10^{-7} M_\sun$, the wind stops.
\par
     The WD photosphere begins to shrink quickly after 
the wind stops.  It takes 34 days to collapse from $0.087~R_\sun$
to $0.03~R_\sun$.  The photospheric temperature increases 
quickly from 30 eV to 40 eV during this period, which is
very consistent with the blackbody temperature of
supersoft X-rays \citep{sch93}.
When the wind stops, the wind mass loss rate decreases 
from $\dot M_{\rm wind}= 1.0 \times 10^{-8} M_\sun$~yr$^{-1}$ to
zero in a day.  This is again 
very consistent with a sharp emergence of supersoft X-rays 
\citep[about a day or so, see][]{rei00}.
\par
     In Figure \ref{binary_flow},  during the strong wind phase 
the schematic mass flow goes to much below the level of full.
After the wind stops,
it gradually goes up and eventually reaches the level of full.
It takes 13.5 days in Figure \ref{vmagfit_rxj0513_self}.
Then, the mass accretion to the WD resumes 20.5 days
after it reached the level of full, i.e., $t_{\rm vis}$ delay time.
It takes more 6 days that the WD envelope expands to blow a wind
again.  The total duration of the low state amounts to
$13.5 + 20.5 + 6 = 40$~days.  Thus, the system repeats 
the cycle mentioned above.
\par
     The modeled system reaches a limit cycle after one cycle of high 
and low states.  The long-term light curve modulations are well
reproduced.  The time-averaged mass transfer rate is 
$1.1 \times 10^{-6} M_\sun$~yr$^{-1}$ and 35\% of the transferred
matter is blown in the wind, so that the MS loses its mass 
at the average rate
of $3.9 \times 10^{-6} M_\sun$~yr$^{-1}$ by stripping effects.
These values satisfy the conditions of $\dot M_{\rm MS}= 
\dot M_{\rm acc} + \dot M_{\rm strip}$ and $\dot M_{\rm strip} =
10 ~ \dot M_{\rm wind}$ in averaged values. 
\par
     It should be noted that our $t_{\rm vis}$ adopted here
is 20.5 days, which is tuned to reproduce the duration of
optical low states, i.e., $t_{\rm LS}= 40$ days, because the exact
number of $t_{\rm vis}$ is still theoretically not known to us.
The duration of low states shortens to $t_{\rm LS}= 25$ days
when $t_{\rm vis}= 18$ days.  Optical low states 
disappear for $t_{\rm vis} \lesssim 14$ days.
On the other hand, it becomes
longer $t_{\rm LS}= 50$ days when $t_{\rm vis}= 23$ days.
We summarize these results in Table \ref{high_low_states}.
The effect of $t_{\rm vis}$ on the long-term light curves 
will be discussed more below in relation to other system 
parameters.
\par
     Thus, we are able to reproduce the basic observational features 
(1)$-$(4) of RX~J0513 summarized in \S 1.  Furthermore, our model
naturally explains two other observational features:
(5) the WD photosphere continues to expand during the rapid mass accretion
phase and still goes on until $\sim 20$ days after the wind started.
This makes a bump of $\sim 0.1$ mag at the head of optical
high states.  (6) The WD photosphere gradually shrinks after the rapid
mass accretion stops.  This also make a gradual decay of $\sim 0.2$ mag
until the end of the wind phase (until the end of the optical high state). 

\placetable{high_low_states}
\placefigure{vmagfit_rxj0513_self_acc}

\subsection{Mass transfer rate of MS}
     The mass transfer rate of the MS companion depends on its
evolutionary stage and binary parameters 
\citep[e.g.,][]{lih97, lan00}.  Since we do not know
the exact number of the mass transfer rate in RX~J0513, 
we have examined various values of the original mass transfer
rate $\dot M_{\rm MS}$, i.e.,
$0.6$, $0.7$, $0.8$, $0.9$, $1.0$, $2.0$, $3.0$, $5.0$ 
(the template), $7.0$, $9.0$, $10.0$, $15.0$, and $20.0$ 
in units of $10^{-6} M_\sun$~yr$^{-1}$,
as summarized in Table \ref{high_low_states}.
The duration of optical high states becomes longer for the higher
mass transfer rates of $\dot M_{\rm MS}$.  On the other hand,
high/low states almost become a 50\% duty cycle
as the mass transfer rate
approaches the critical value, i.e., $\dot M_{\rm MS} \rightarrow 
\dot M_{\rm cr}$. 
Some of the calculated light curves are shown
in Figure \ref{vmagfit_rxj0513_self_acc}.
\par
     A remarkable result is that the durations of low states
are kept to be $\sim 40$ days for a wide range of the mass
transfer rates, i.e.,
$7 \times 10^{-7} M_\sun$~yr$^{-1}
\lesssim \dot M_{\rm MS} \lesssim 6 \times 10^{-6} M_\sun$~yr$^{-1}$.
The duration of high states varies largely from
40 days to 120 days for the same range of the mass transfer rates.
This is just the result that we want to explain theoretically,
because it is an observational feature as shown in Figure
\ref{vmagfit_rxj0513_self} (top panel).
The low state gradually shortens and eventually disappears 
as the mass transfer rate increases to 
$\dot M_{\rm MS} \gtrsim 1.5 \times 10^{-5} M_\sun$~yr$^{-1}$
for a set of $t_{\rm vis}=20.5~$days and $c_1=10.0$.  
The luminosity of high states is brighter
for higher mass transfer rates of $\dot M_{\rm MS}$.

\placefigure{vmagfit_rxj0513_self_wdm1350}
\placefigure{vmagfit_rxj0513_self_wdm1200}
\placefigure{vmagfit_rxj0513_self_wdm1100}

\subsection{White dwarf mass}
\begin{figure}
\plotone{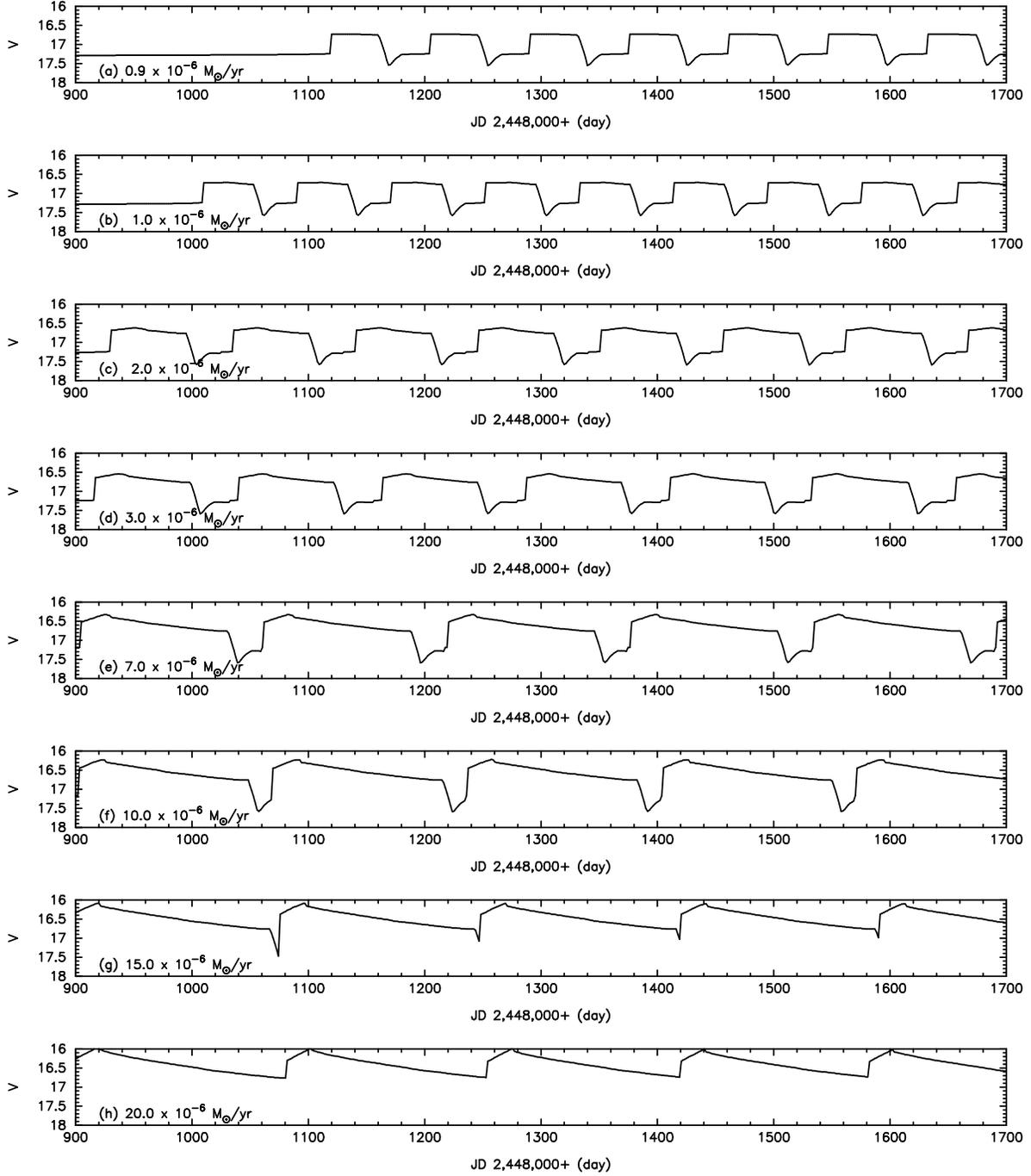}
\caption{
Calculated $V$-magnitude light curves are plotted against time
(JD 2,448,000+) for various mass transfer rates.
The model parameters for each case are
also summarized in Table \ref{high_low_states}:
(a) $\dot M_{\rm MS}= 0.9 \times 10^{-6}M_\sun$~yr$^{-1}$.
It takes about 210 days  for this relatively low mass 
accretion rate that the WD envelope mass reaches
$6.5 \times 10^{-7} M_\sun$, where the WD begins to blow a wind,
from the initial envelope mass of $5.0 \times 10^{-7} M_\sun$.
The first flat part corresponds to a part of this 210 days.  
(b) $\dot M_{\rm MS}= 1.0 \times 10^{-6}M_\sun$~yr$^{-1}$,
(c) $\dot M_{\rm MS}= 2.0 \times 10^{-6}M_\sun$~yr$^{-1}$,
(d) $\dot M_{\rm MS}= 3.0 \times 10^{-6}M_\sun$~yr$^{-1}$,
(e) $\dot M_{\rm MS}= 7.0 \times 10^{-6}M_\sun$~yr$^{-1}$,
(f) $\dot M_{\rm MS}= 10. \times 10^{-6}M_\sun$~yr$^{-1}$,
(g) $\dot M_{\rm MS}= 15. \times 10^{-6}M_\sun$~yr$^{-1}$,
(h) $\dot M_{\rm MS}= 20. \times 10^{-6}M_\sun$~yr$^{-1}$.
Each model has $M_{\rm WD}= 1.3~M_\sun$, $c_1=10.0$, 
and $t_{\rm vis}= 20.5$ days in common.  We adopt the apparent
distance modulus of $(m-M)_V = 18.7$. 
\label{vmagfit_rxj0513_self_acc}}
\end{figure}

\begin{figure}
\plotone{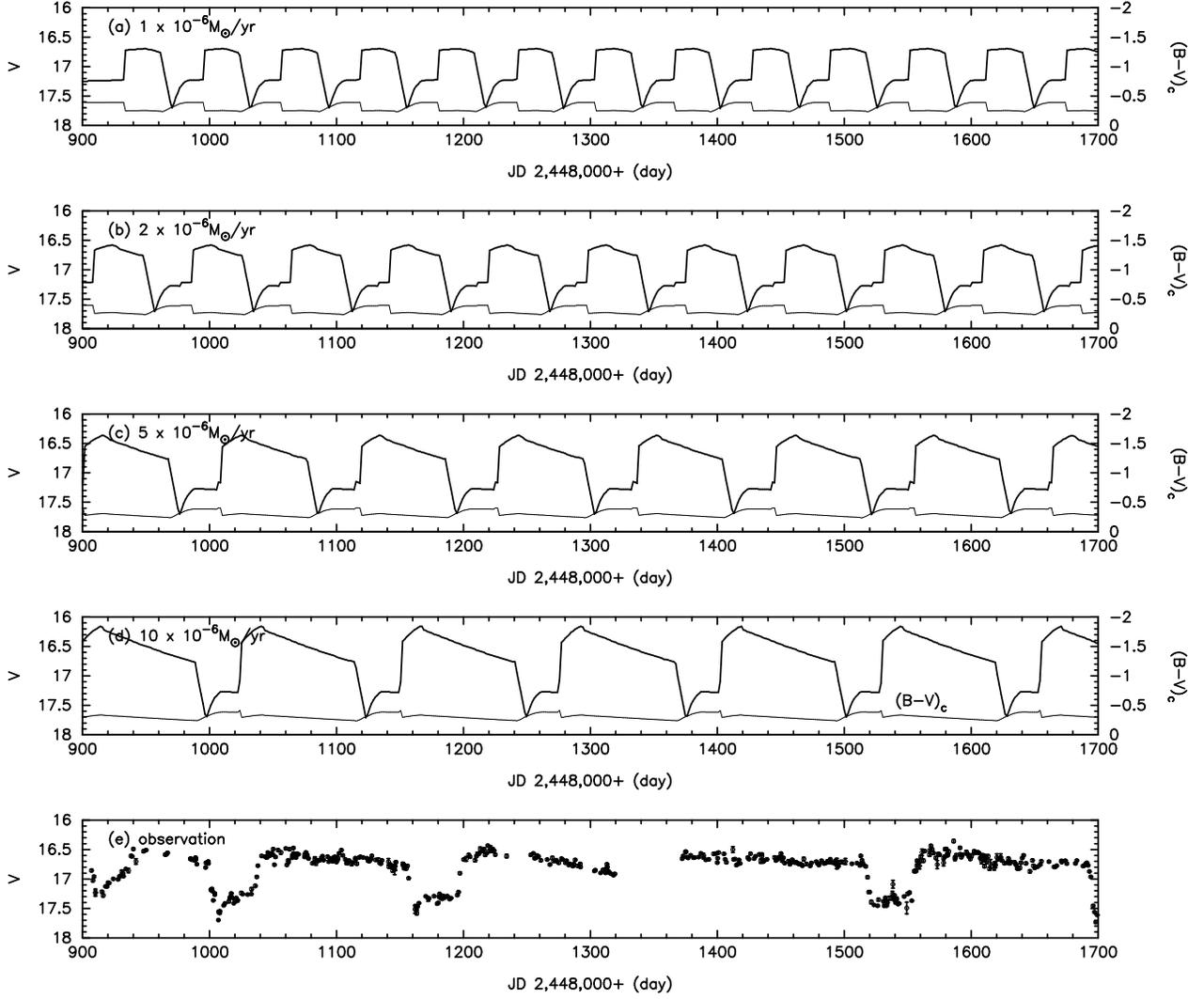}
\caption{
Calculated $V$-magnitude and $B-V$ color light curves are plotted
against time (JD 2,448,000+) for our $M_{\rm WD}= 1.35~M_\sun$
models.  The model parameters are summarized in 
Table \ref{high_low_states}:
(a) $\dot M_{\rm MS}= 1.0 \times 10^{-6}M_\sun$~yr$^{-1}$,
(b) $\dot M_{\rm MS}= 2.0 \times 10^{-6}M_\sun$~yr$^{-1}$,
(c) $\dot M_{\rm MS}= 5.0 \times 10^{-6}M_\sun$~yr$^{-1}$,
(d) $\dot M_{\rm MS}= 10.0 \times 10^{-6}M_\sun$~yr$^{-1}$, and
(e) observational $V$-magnitudes are also added 
\citep[taken from][]{alc96}.
The other parameters of $c_1= 10.0$ and $t_{\rm vis}= 15.0$ days
are common among the model light curves.  We adopt the apparent
distance modulus of $(m-M)_V = 18.7$.  
\label{vmagfit_rxj0513_self_wdm1350}}
\end{figure}

\begin{figure}
\plotone{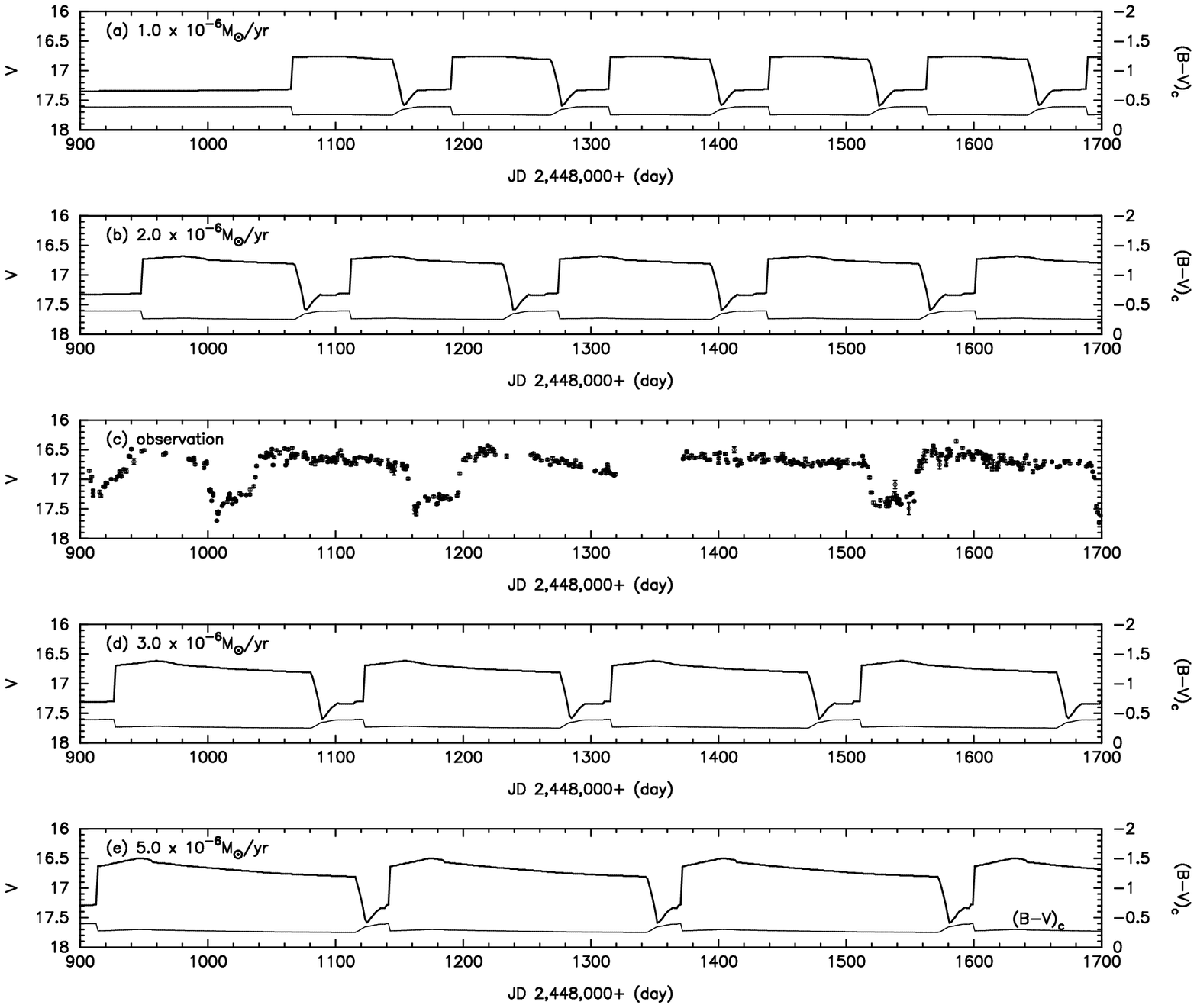}
\caption{
Calculated $V$-magnitude and $B-V$ color light curves are plotted
against time (JD 2,448,000+) for our $M_{\rm WD}= 1.2~M_\sun$
models.  The model parameters are summarized in 
Table \ref{high_low_states}:
(a) $\dot M_{\rm MS}= 1.0 \times 10^{-6}M_\sun$~yr$^{-1}$,
(b) $\dot M_{\rm MS}= 2.0 \times 10^{-6}M_\sun$~yr$^{-1}$,
(c) observational $V$-magnitudes \citep[taken from][]{alc96},
(d) $\dot M_{\rm MS}= 3.0 \times 10^{-6}M_\sun$~yr$^{-1}$,
(e) $\dot M_{\rm MS}= 5.0 \times 10^{-6}M_\sun$~yr$^{-1}$.
The other parameters of $c_1= 10.0$ and $t_{\rm vis}= 32.0$ days
are common among the model light curves.  We adopt the apparent
distance modulus of $(m-M)_V = 18.7$. 
\label{vmagfit_rxj0513_self_wdm1200}}
\end{figure}

\begin{figure}
\plotone{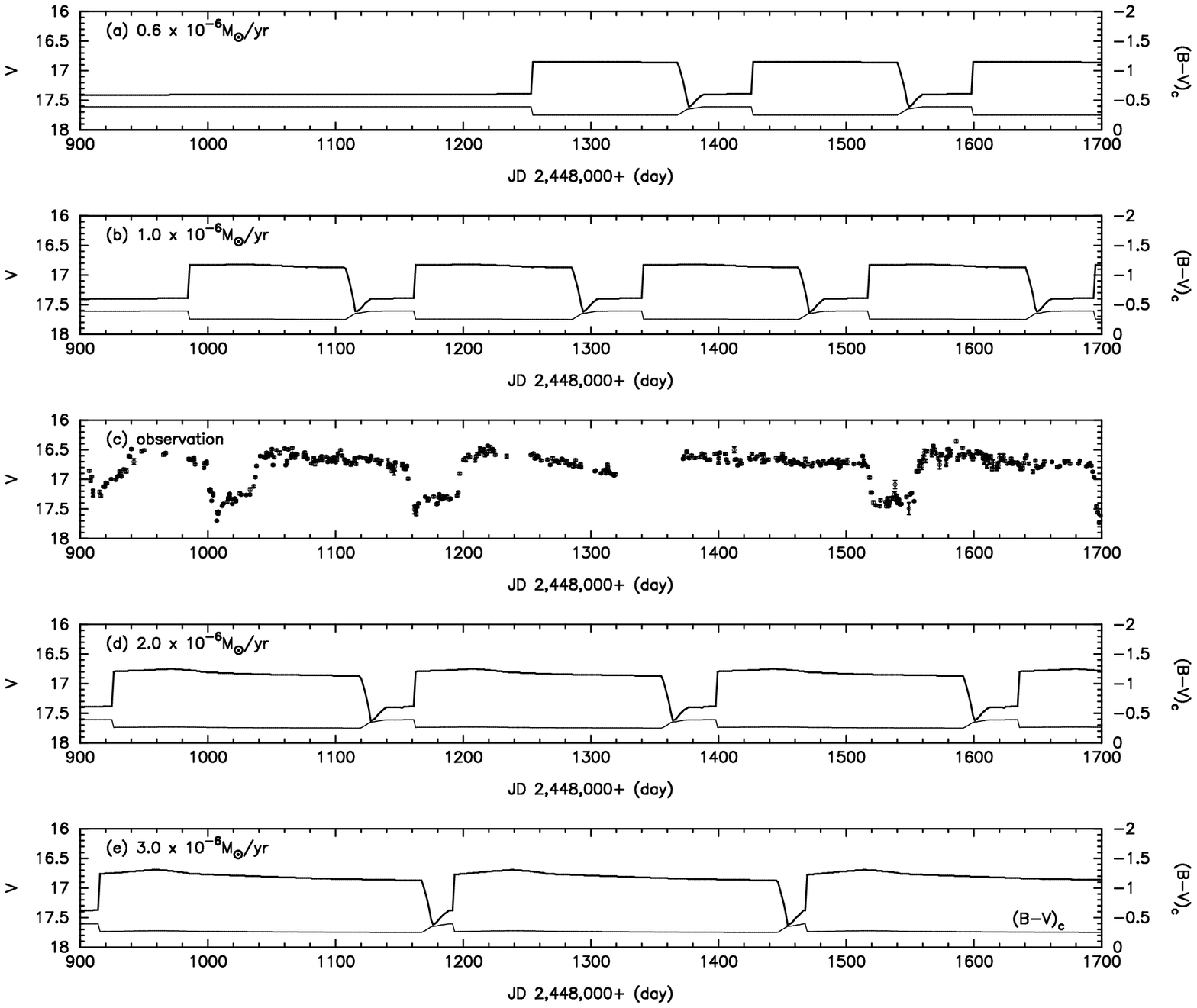}
\caption{
Calculated $V$-magnitude and $B-V$ color light curves are plotted
against time (JD 2,448,000+) for our $M_{\rm WD}= 1.1~M_\sun$
models.  The model parameters are summarized in 
Table \ref{high_low_states}:
(a) $\dot M_{\rm MS}= 0.6 \times 10^{-6}M_\sun$~yr$^{-1}$,
(b) $\dot M_{\rm MS}= 1.0 \times 10^{-6}M_\sun$~yr$^{-1}$,
(c) observational $V$-magnitudes \citep[taken from][]{alc96},
(d) $\dot M_{\rm MS}= 2.0 \times 10^{-6}M_\sun$~yr$^{-1}$,
(e) $\dot M_{\rm MS}= 3.0 \times 10^{-6}M_\sun$~yr$^{-1}$.
The other parameters of $c_1= 10.0$ and $t_{\rm vis}= 45.0$ days
are common among the model light curves.  We adopt the apparent
distance modulus of $(m-M)_V = 18.7$. 
\label{vmagfit_rxj0513_self_wdm1100}}
\end{figure}

     We have also examined various WD masses,
i.e., $M_{\rm WD}= 1.35$, $1.2$, and $1.1~M_\sun$,
with the mass ratio being kept $q= M_{\rm MS} / M_{\rm WD} = 2$.
The results are shown in Figures  
\ref{vmagfit_rxj0513_self_wdm1350}$-$
\ref{vmagfit_rxj0513_self_wdm1100}.
In general, more massive white dwarfs have a larger wind mass loss
rate, so that the duration of the wind phase becomes much shorter.
Because we do not know the exact number of $t_{\rm vis}$,
our guided principle is that we tune $t_{\rm vis}$ to satisfy 
the duration of the low state is 40 days, i.e., $t_{\rm LS}= 40$ days.
For the case of $M_{\rm WD}= 1.35~M_\sun$ shown in Figure
\ref{vmagfit_rxj0513_self_wdm1350}, we have tuned the viscous timescale
$t_{\rm vis}$ 
at the mass transfer rate of $\dot M_{\rm MS}= 5 \times 10^{-6}
M_\sun$~yr$^{-1}$.  The tuned viscous timescale is 
$t_{\rm vis}= 15.0$ days.  The durations of high/low states
are similar to those for $M_{\rm WD}= 1.3~M_\sun$ in Figure
\ref{vmagfit_rxj0513_self_acc} except for the relatively short
durations.  The WD mass in RX~J0513 may be less massive than 
$1.35~M_\sun$, because the gradual decay of the light curve 
in high states seems to be too fast to be compatible
with the observation.
\par
     For the case of $M_{\rm WD}= 1.2~M_\sun$ shown in Figure
\ref{vmagfit_rxj0513_self_wdm1200}, we have tuned the viscous timescale
at the mass transfer rate of $\dot M_{\rm MS}= 3 \times 10^{-6}
M_\sun$~yr$^{-1}$,
i.e., $t_{\rm vis}= 32.0$ days.  The durations of high states
are longer than those for $M_{\rm WD}= 1.3~M_\sun$
in Figure \ref{vmagfit_rxj0513_self_acc}.
The WD mass in RX~J0513 may be somewhere between $1.2~M_\sun$ and
$1.3~M_\sun$, because the decay of the light curve in high states
seems to be slightly slower than, but in reasonable agreement with, 
the observation.  
\par
     For the case of $M_{\rm WD}= 1.1~M_\sun$ shown in Figure
\ref{vmagfit_rxj0513_self_wdm1100}, we have tuned the viscous timescale
at the mass transfer rate of $\dot M_{\rm MS}= 2 \times 10^{-6}
M_\sun$~yr$^{-1}$, i.e., $t_{\rm vis}= 45.0$ days.
The durations of high states are much longer than those 
for $M_{\rm WD}= 1.3~M_\sun$ in Figure
\ref{vmagfit_rxj0513_self_acc}.
The WD mass of $1.1~M_\sun$ may not be the case,
because the decay of the light curve in high states is too slow
to be compatible with the observation.

\begin{figure}
\plotone{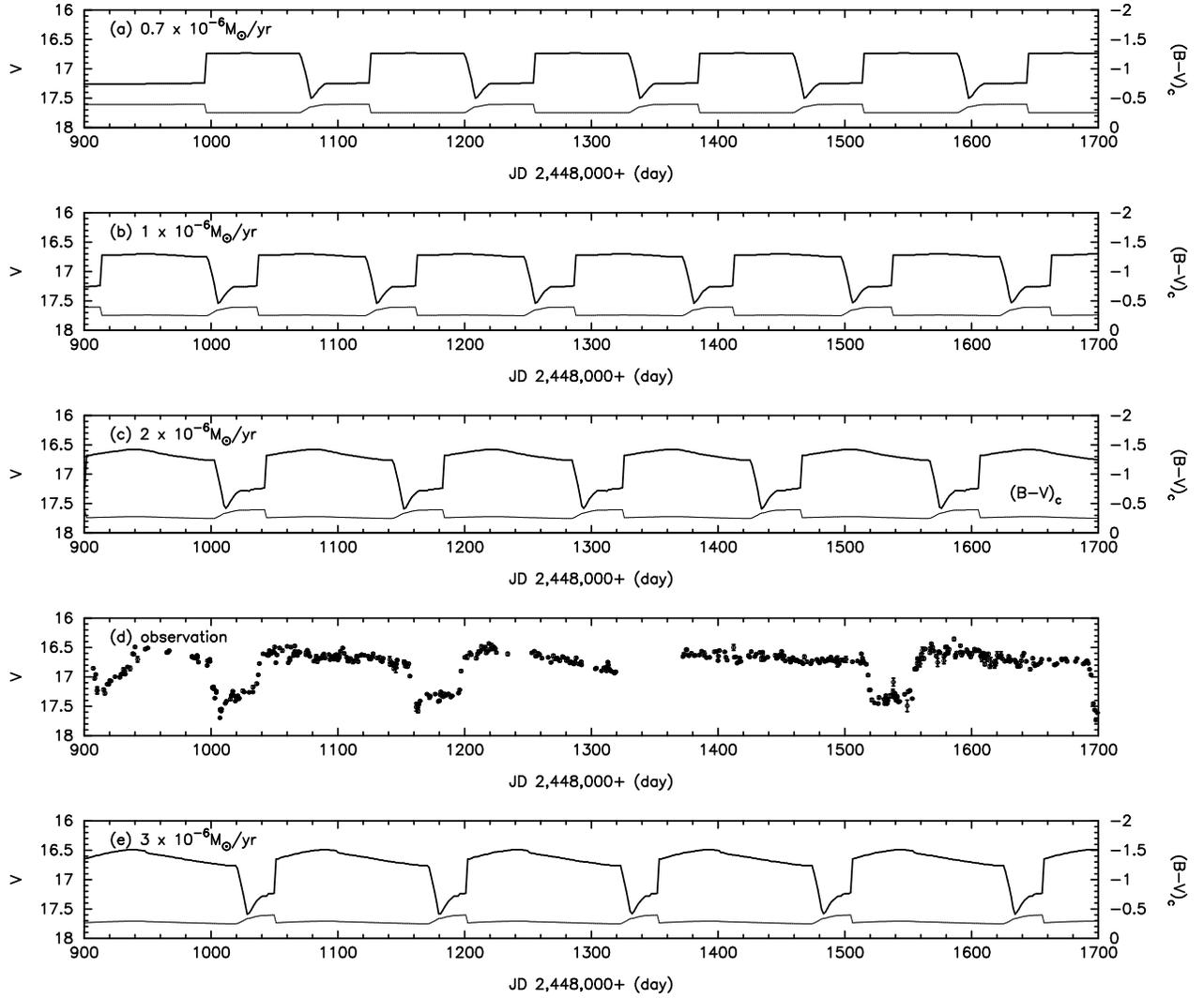}
\caption{
Calculated $V$-magnitude and $B-V$ color light curves are plotted
against time (JD 2,448,000+) for a different value of 
$c_1= 5.0$.  The model parameters are summarized in 
Table \ref{high_low_states}:
(a) $\dot M_{\rm MS}= 0.7 \times 10^{-6}M_\sun$~yr$^{-1}$,
(b) $\dot M_{\rm MS}= 1.0 \times 10^{-6}M_\sun$~yr$^{-1}$,
(c) $\dot M_{\rm MS}= 2.0 \times 10^{-6}M_\sun$~yr$^{-1}$,
(d) observational $V$-magnitudes \citep[taken from][]{alc96},
(e) $\dot M_{\rm MS}= 3.0 \times 10^{-6}M_\sun$~yr$^{-1}$.
The other parameters of $M_{\rm WD}= 1.3~M_\sun$ 
and $t_{\rm vis}= 36.0$ days
are common among the model light curves.  We adopt the apparent
distance modulus of $(m-M)_V = 18.7$. 
\label{vmagfit_rxj0513_self_wdm1300_strip_c5}}
\end{figure}

\begin{figure}
\plotone{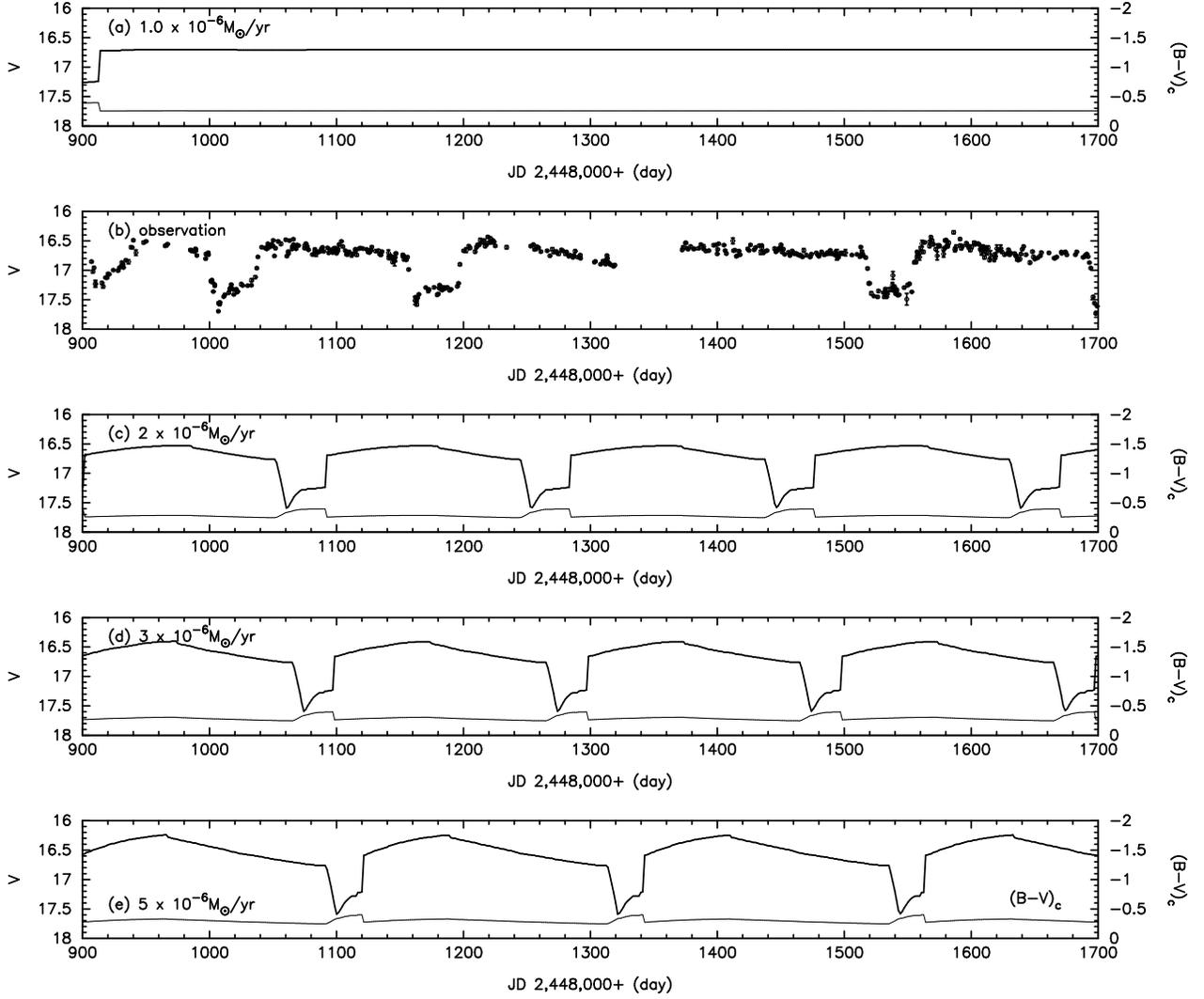}
\caption{
Calculated $V$-magnitude and $B-V$ color light curves are plotted
against time (JD 2,448,000+) for a different value of 
$c_1= 1.5$.  The model parameters are summarized in 
Table \ref{high_low_states}:
(a) $\dot M_{\rm MS}= 1.0 \times 10^{-6}M_\sun$~yr$^{-1}$,
(b) observational $V$-magnitudes \citep[taken from][]{alc96},
(c) $\dot M_{\rm MS}= 2.0 \times 10^{-6}M_\sun$~yr$^{-1}$,
(d) $\dot M_{\rm MS}= 3.0 \times 10^{-6}M_\sun$~yr$^{-1}$,
(e) $\dot M_{\rm MS}= 5.0 \times 10^{-6}M_\sun$~yr$^{-1}$.
The other parameters of $M_{\rm WD}= 1.3~M_\sun$ 
and $t_{\rm vis}= 62.0$ days
are common among the model light curves.  We adopt the apparent
distance modulus of $(m-M)_V = 18.7$. 
\label{vmagfit_rxj0513_self_wdm1300_strip_c015}}
\end{figure}

\placefigure{vmagfit_rxj0513_self_wdm1300_strip_c5}
\placefigure{vmagfit_rxj0513_self_wdm1300_strip_c015}

\subsection{Mass stripping effect}
      It is a difficult work to accurately estimate the effect of
mass-stripping, that is, to determine the coefficient of $c_1$ in
equation (\ref{mass_stripping_rate}).  Therefore, we calculate
two other cases of $c_1$, i.e., $c_1=5$ and $c_1=1.5$ 
and examine how the light curves depend on the coefficient.
It is obvious that strong modulation of the mass transfer rate
never occurs for the case of $c_1 \le 1$ because there is no case of 
$\dot M_{\rm MS} < \dot M_{\rm strip} = c_1 \dot M_{\rm wind}
(\le \dot M_{\rm wind} < \dot M_{\rm acc} < \dot M_{\rm MS})$.
Our calculated models and their parameters are listed 
in Table \ref{high_low_states}.
\par
     For the case of $c_1 = 5$ shown in Figure 
\ref{vmagfit_rxj0513_self_wdm1300_strip_c5},
the overall features of light curves
are essentially similar to the case of $c_1 = 10$ shown in
Figure \ref{vmagfit_rxj0513_self_acc}.  Main differences are
that (1) the tuned viscous timescale is $t_{\rm vis}= 36.0$ days 
at $\dot M_{\rm MS}= 2 \times 10^{-6} M_\sun$~yr$^{-1}$, 
much longer than $t_{\rm vis}= 20.5$ days for $c_1 = 10$,
and (2) bumps at the head of high states continue for a much longer
period and may not be compatible with the observation.
\par
     For the case of $c_1 = 1.5$ shown in Figure
\ref{vmagfit_rxj0513_self_wdm1300_strip_c015}, 
a new thing is the appearance of no modulation in mass transfer
for a range of 
$0.67 \times 10^{-6} M_\sun$~yr$^{-1} < \dot M_{\rm MS} < 
1.5 \times 10^{-6} M_\sun$~yr$^{-1}$.
The tuned viscous timescale is $t_{\rm vis}= 62.0$ days 
at $\dot M_{\rm MS}= 2 \times 10^{-6} M_\sun$~yr$^{-1}$. 
Bumps at the head of high states continue for about 40 days.
Both the viscous timescale and the duration of bumps are
much longer than those for $c_1= 10$ and $c_1= 5$.
The long duration of bumps is not compatible with the observation.
\par
     An important result in this subsection is that modulation
in mass transfer occurs and optical high/low states appear even
for relatively low efficiencies of mass-stripping on the MS, 
i.e., for $c_1 = 1.5$, if the viscous timescale is relatively
long, i.e., $t_{\rm vis}= 62$ days.
Such a low case of $c_1$ may correspond to a low velocity case
of winds as understood from 
equation (\ref{mass_stripping_rate_real}) and a relatively long
viscous timescale is realized for binaries with a longer orbital
period, for example, longer than a day, as derived from 
equation (\ref{viscous_timescale}).

\placefigure{thermal_time}
\placefigure{vmagfit_rxj0513_self_wdm1300_angle}
\placefigure{v_mag_wd13ms26_wind_high_orbital}

\section{DISCUSSION}
     It has been argued that only a viable mechanism 
for the transition between the optical
high/X-ray off and optical low/X-ray on states is the
contraction/expansion of the WD photosphere resulting from
the variation of the mass accretion rate to the WD 
\citep[e.g.,][]{mey97, pak93, rei96, rei00, sou96}, although a few
other mechanisms have ever been proposed \citep[see, e.g., 
discussion of][]{sou96}.  Therefore, we first examine whether or
not this mechanism can explain all the observational features.
\par
\begin{figure}
\plotone{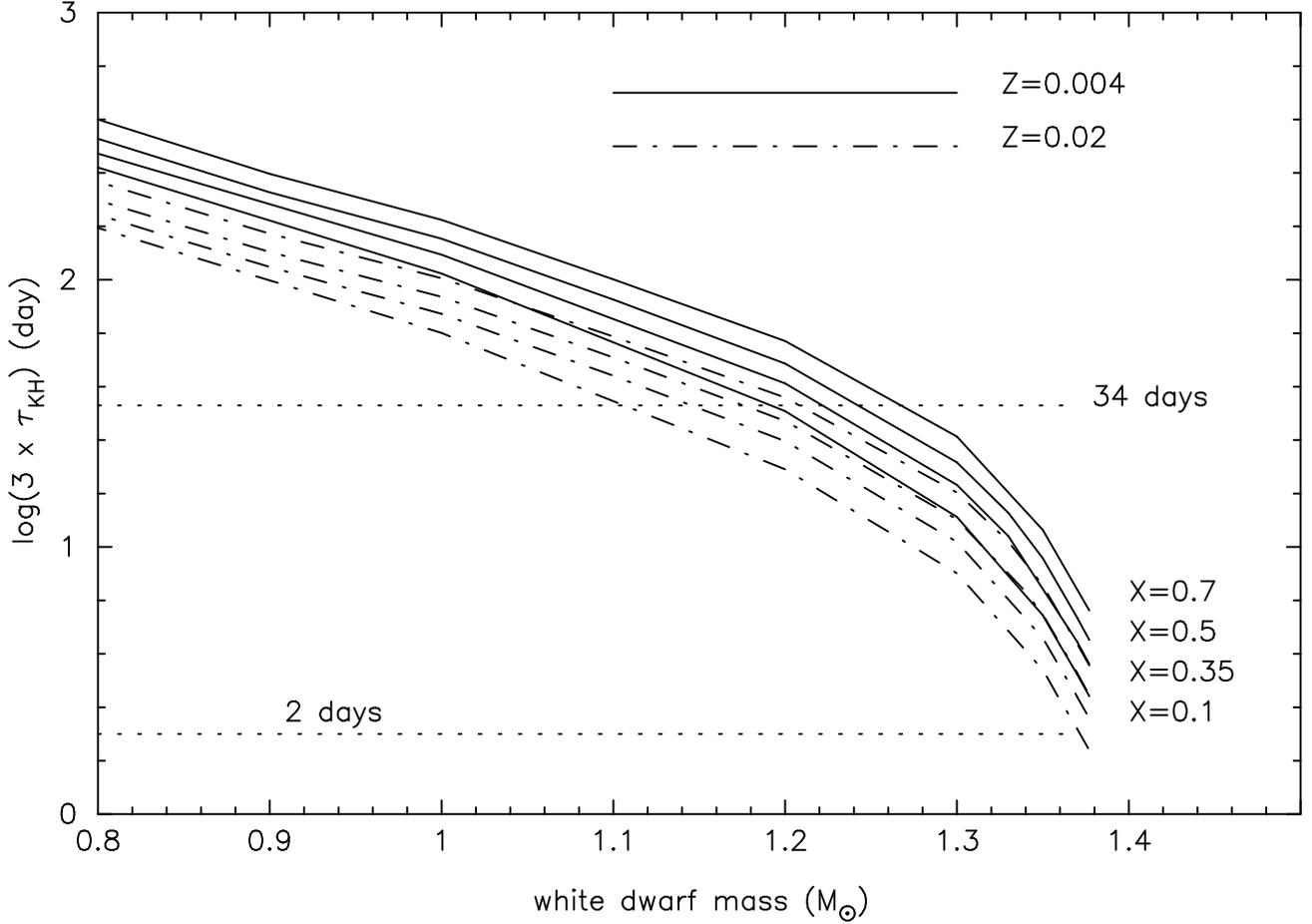}
\caption{
Three times the Kelvin-Helmholtz timescale of the WD envelope 
is plotted against the WD mass.
Since the timescale depends on 
the metallicity ($Z$) and the hydrogen content ($X$), we plot
four cases of the hydrogen content, $X=0.1$, 0.35, 0.5, and 0.7,
number of which is attached to each line, each for two cases of 
the metallicity, $Z=0.004$ ({\it solid line}) 
and $Z=0.02$ ({\it dashed-dotted line}).  
Here we have calculated $\tau_{\rm KH} =$(thermal energy of the WD 
envelope)$/$(luminosity of the WD) just after the wind stops.
This corresponds to the shortest timescale that the WD photosphere 
shrinks by a factor of four, i.e.,
from $\sim 0.08~R_\sun$ ($\sim 25$ eV) 
to $\sim 0.02~R_\sun$ ($\sim 50$ eV).
\label{thermal_time}}
\end{figure}
\begin{figure}
\plotone{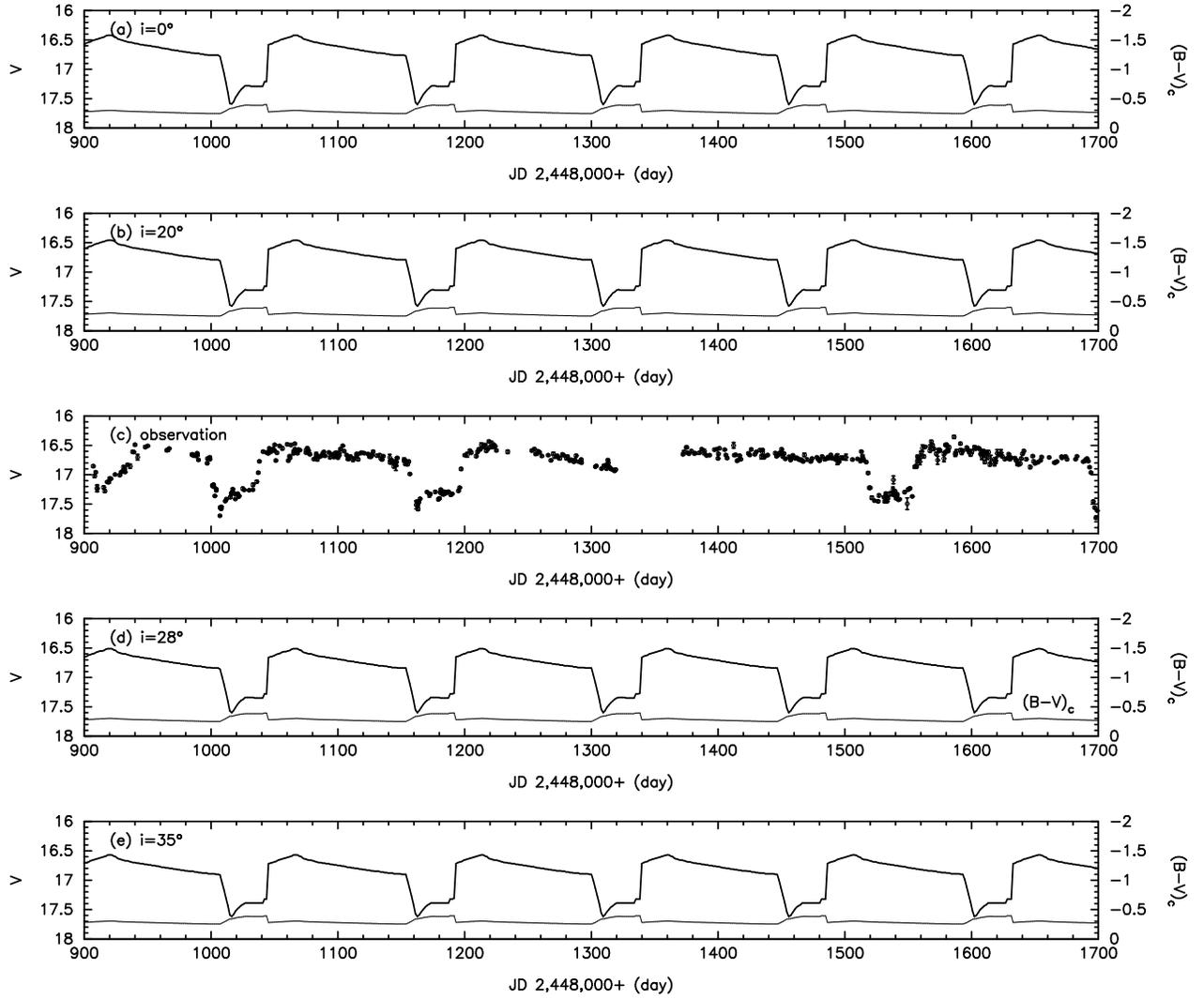}
\caption{
Calculated $V$-magnitude and $B-V$ color light curves are plotted
against time (JD 2,448,000+) for various inclination angles
of (a) $i=0\arcdeg$, (b) $i=20\arcdeg$, (c) MACHO observation
taken from \citet{alc96}, (d) $i=28\arcdeg$, 
and (d) $i=35\arcdeg$.
The other model parameters are the same as those 
in Fig. \ref{vmagfit_rxj0513_self}.
\label{vmagfit_rxj0513_self_wdm1300_angle}}
\end{figure}
\begin{figure}
\plotone{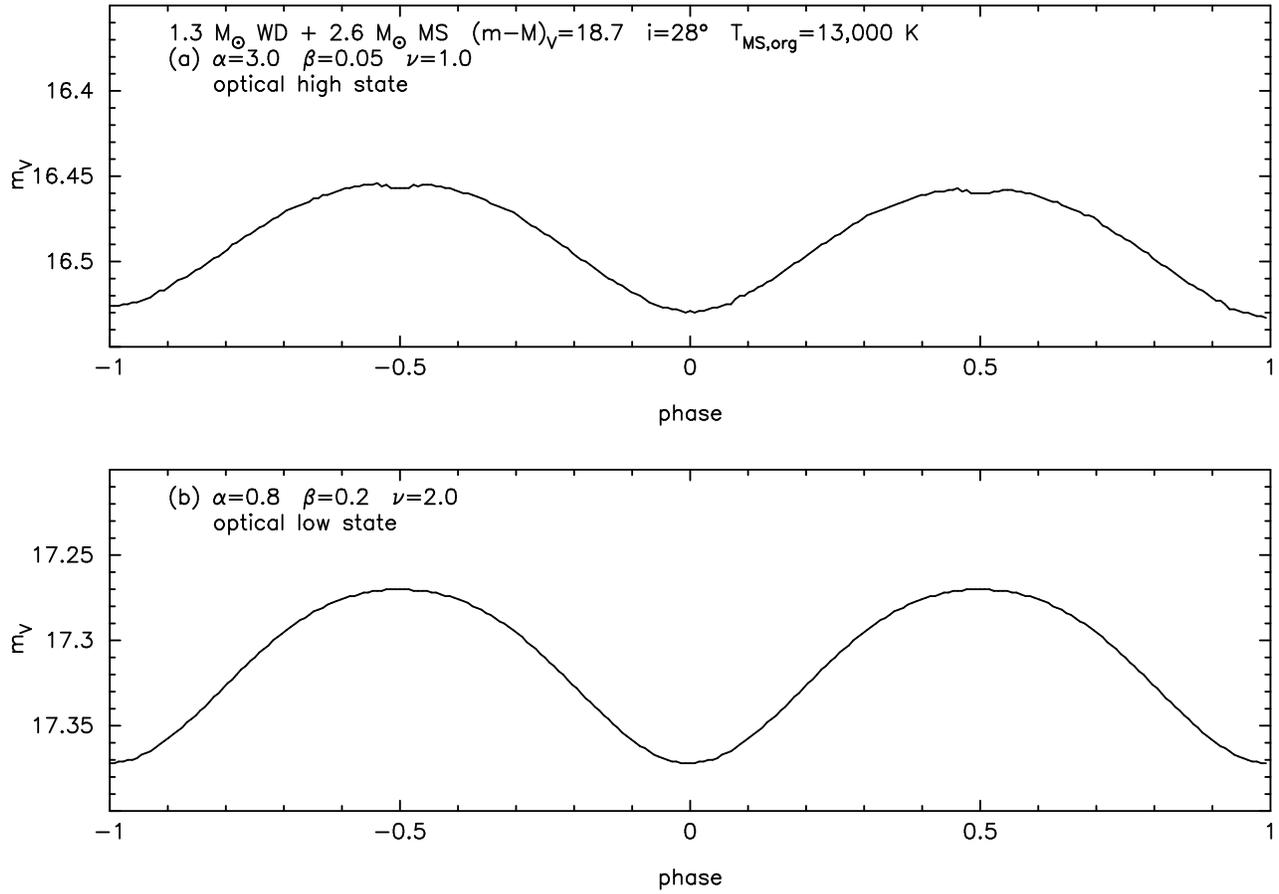}
\caption{
Calculated $V$-magnitudes are plotted against two orbital phases
(from $-1.0$ to $1.0$) for (a) the optical
high state and (b) the optical low state.
We adopt the inclination angle of $i=28\arcdeg$.
The other model parameters are the same as those 
in Fig. \ref{vmagfit_rxj0513_self}.
\label{v_mag_wd13ms26_wind_high_orbital}}
\end{figure}

     \citet{rei00} detected a very sharp rise ($< 3$ days)
and a steep decline ($< 2$ days) in supersoft X-rays by a factor
of $> 100$.  This requires a contraction/expansion of the WD
photosphere by a factor of $> 7$ and an increase/decrease of the
photospheric temperature by a factor of $> 3$ \citep{rei00}.
We have estimated the timescale of contracting/expanding 
WD envelopes based on our wind and static solutions.
The WD photosphere gradually shrinks during the wind phase while
it begins to contract most rapidly just after the wind stops 
(see the bottom panel of Fig. \ref{vmagfit_rxj0513_self}).
Even for this most rapid shrink phase, it takes about
three Kelvin-Helmholtz timescale to shrink by a factor of four, i.e., 
\begin{equation}
t_{\rm contract} \sim 3 \tau_{\rm KH}, 
\end{equation}
from our model calculations.
This Kelvin-Helmholtz timescale is defined by
\begin{equation}
\tau_{\rm KH}= {{\Delta E_{\rm env, thermal}} \over {L}},
\end{equation}
where $\Delta E_{\rm env,~thermal}$ is the thermal energy
of the WD envelope, and $L$ the luminosity of the WD, just when
the wind stops.  These contraction timescales just after the wind stops 
are plotted against various WD masses in Figure \ref{thermal_time}.
It is obvious that the rapid ($< 2$ days) change in the soft X-ray
flux cannot be explained by the contraction
of the WD photosphere even if we assume the extreme case
of $M_{\rm WD}= 1.377~M_\sun$ (at least, for $Z=0.004$).
\par
     For the expansion case of the WD photosphere during the rapid
mass accretion phase, our steady-state wind model does not 
correctly include the compression term by accretion so that 
the estimated timescale is not so accurate.  In this sense,
our numerical results cannot give a strong constraint on the timescale
of photospheric expansion.  It should be pointed out, however, that
usual time-dependent Henyey methods 
fail to be converged in a rapidly expanding photosphere of WDs
and are now unable to involve wind mass loss.
\par
     \citet{rei00} suggested a characteristic decay timescale of 
34 days from the exponential decay of supersoft X-rays in an optical
low state.  In our model, there are two possibilities.  One is 
the timescale of expansion/contraction of the WD photosphere
and the other is the viscous timescale of the accretion disk, 
$t_{\rm vis}$.  If this 34 days timescale is related to
the expansion/contraction timescale of a WD envelope,
its WD mass is somewhere between $1.2 ~M_\sun$ and $1.3 ~M_\sun$
for the metallicity of $Z=0.004$ as shown 
in Figure \ref{thermal_time}.  This mass range of the WD is
consistent with our long-term light curve modulations.
On the other hand, if it indicates that $t_{\rm vis}= 34$ days,
the ranges of the parameters are also suggested to be 
$M_{\rm WD}= 1.2 - 1.3 ~M_\sun$ and $c_1= 5-10$ from 
Table \ref{high_low_states}.  These are also consistent with
our long-term light curve modulations.
\par
     \citet{sou96} suggested that star spots on the secondary 
surface cover the L1 point, resulting in a decreased mass 
transfer rate from the secondary.  This mechanism has been
originally suggested by \citet{liv94} for the VY Scl stars.
In RX~J0513, however, it is very unlikely that the MS companion 
develops star spots on its surface, because its mass is 
suggested to be $2-3 ~M_\sun$, not having a deep surface convection.
\par
      Very recently, \citet{hut02} reported radial velocity modulations
of $\sim 117 \pm 40$ and $54 \pm 10$ km~s$^{-1}$ for the broad 
\ion{O}{6} emission and Lyman absorption, respectively.
If these trace the motions of the hot and cool components,
respectively, the estimated mass ratio is 
$q = M_{\rm MS} / M_{\rm WD} = 2.2 \pm 1.2$,
being consistent with our assumption of $q=2$.
The corresponding inclination angle is 
$i= 28\arcdeg \pm 10 \arcdeg$ for our case of $M_{\rm WD} = 
1.3 ~M_\sun$ and $M_{\rm MS} = 2.6 ~M_\sun$, which is a bit
larger than our assumption of $i= 10\arcdeg$.  Then,
we have examined how the long-term light curve depends on 
the inclination.  Additional four cases of the inclination angle,
i.e., $i=0\arcdeg$,  $20\arcdeg$, $28\arcdeg$,  and $35\arcdeg$, 
are calculated.  The results are also shown
in Figure \ref{vmagfit_rxj0513_self_wdm1300_angle}, which are
not so much different in shape from the case of $i=10\arcdeg$,
but the brightness in optical high states decreases by 
a factor of $\cos i$.
\par
      \citet{cow02} have reanalyzed the MACHO data for eight
years and determined a new orbital period and a new ephemeris.
They found that the orbital modulations of the optical light
curve are different in shape between the optical high and low states.
The orbital modulation dip is much broader in the high state than
in the low state.  Assuming an inclination angle of $i=28\arcdeg$, 
we have calculated two orbital modulations 
each for the optical high state and for the optical low state
as shown in Figure \ref{v_mag_wd13ms26_wind_high_orbital}.
The observational broad and shallow orbital modulations in 
the optical high state are reasonably reproduced as well as their full
extents (amplitudes) of $\Delta V \sim 0.06$ mag.
On the other hand, the full extents of the orbital modulations
become larger (deeper) to $\Delta V \sim 0.09$ mag in the optical 
low state, being consistent with the observation.

\section{CONCLUSIONS}
     A new self-sustained model for the long-term light curve variations
of \objectname{RX~J0513.9$-$6951} is formulated based 
on our optically thick wind
model of mass-accreting white dwarfs.  We also show that 
\objectname{RX~J0513.9$-$6951} is the first beautiful example of the 
accretion wind evolution, which is a key evolutionary process 
to Type Ia supernovae \citep[e.g.,][]{hac02, hac01kb}
\par
     We have already shown that, when the mass accretion rate to a WD
exceeds the critical rate $\dot M_{\rm cr} \approx 0.75 \times 10^{-6}
(M_{\rm WD}/M_\sun - 0.4) M_\sun$~yr$^{-1}$, 
optically thick strong winds begin to blow from the WD.  
As a result, a formation of common envelope is avoided.
In such a situation, the WD accretes and
burns hydrogen atop the WD at the critical rate.
The excess matter transferred to the WD above the critical rate is
expelled by winds \citep{hkn96, hkn99, hknu99}.  
This is called the accretion wind evolution \citep{hac01kb}.
Unfortunately, no such systems have ever been discovered.
Here, we report that the LMC supersoft X-ray source 
\objectname{RX~J0513.9$-$6951}
is the first example of such an object, i.e., just in the accretion
wind evolution. 
\par
     We summarize our main results as follows:
\par\noindent
{\bf 1.} Mass ejection in the accretion wind evolution
does occur not continuously but intermittently
because the mass transfer is attenuated by strong winds.
The strong winds collide with the secondary and
strip off the very surface layer of the secondary.
Properly formulating this mass stripping effect and the ensuing
decay of mass transfer rate, we obtain, in a
self-sustained manner, on/off of mass transfer from the companion
star. \\
{\bf 2.} A light curve modeling on the 2000 outburst of the 
recurrent nova \objectname{CI Aql} showed that the size of 
the disk around the white dwarf
extends widely up to the companion star or over only during
the strong wind phase \citep{hac03k}.
Including this large extension of a disk in a strong wind phase,
we are able to reproduce the 
transition between the optical high and low states of the LMC 
supersoft X-ray source \objectname{RX~J0513.9$-$6951}:
When a strong wind occurs,
the very surface layer of the disk is blown in the wind and 
the disk expands up to the companion or 
over.  This large extension of the irradiation area makes 
a sharp $\sim 1$ mag rise in the optical light curve.
On the other hand, it drops sharply by 
a magnitude when the wind stops because the size of the disk returns
to its original size. \\
{\bf 3.} In optically thick winds, the mass loss rate rises
to $\dot M_{\rm wind} \gtrsim 1 \times 10^{-8} M_\sun$~yr$^{-1}$
one day after the wind starts.  This can obscure completely
soft X-rays.  On the other hand, the wind mass loss rate drops from 
$\dot M_{\rm wind} \sim 1 \times 10^{-8} M_\sun$~yr$^{-1}$ to zero
in within a day, which enables a short ($< 1$ day) timescale of
soft X-ray emergence.  Thus, we are also able to reproduce a very 
rapid emergence/decay of supersoft X-rays in within a day or so. \\
{\bf 4.} We abstract two important system parameters.  
One is the viscous timescale of the accretion disk, $t_{\rm vis}$,
and the other is the efficiency of mass-stripping by winds, 
$c_1$, defined by $\dot M_{\rm strip} = c_1 \dot M_{\rm wind}$.
Both of them should be determined self-consistently but are not
accurately determined at least at the present time like 
the $\alpha$-parameter of \citet{sha73}.  
We are able to reproduce the optical high/low states
for reasonable ranges of the system parameters, i.e.,
20 days $\lesssim t_{\rm vis} \lesssim$ 60 days and
$1.5 \lesssim c_1 \lesssim 10$. \\
{\bf 5.} The original mass transfer rate from the main-sequence
companion is probably constant in time.  Even if this original 
mass transfer rate varies from several times $10^{-7} M_\sun$~yr$^{-1}$
to several times $10^{-6} M_\sun$~yr$^{-1}$, by a factor of ten,
the duration of optical low states hardly changes, i.e., stays around
$\sim 40$ days, for a fixed set of $(t_{\rm vis}, c_1)$. 
On the other hand, the duration of optical high states varies
from $\sim 40$ days to $\sim 120$ days for the same range
of the original mass transfer rate.  These durations of optical high/low
states are the prominent features of the MACHO observations 
\citep{alc96, cow02}. \\
{\bf 6.} The white dwarf photosphere expands and blows a wind 
during the rapid mass accretion phase.  The photospheric expansion
still goes on for a while after the wind suppresses the mass transfer,
because the accretion disk supplies mass to the WD
for a draining time (viscous timescale).  
This makes a bump of $\sim 0.1$ mag at the head of the optical high state.  
The duration of this bump may determine the value of $t_{\rm vis}$.
The WD photosphere gradually shrinks after the rapid
mass accretion stops.  This makes a gradual decay of $\sim 0.2$ mag
in the optical light curve until the end of the wind phase
(until the end of the optical high state). 
These two light curve features clearly appear 
in the MACHO observation \citep{alc96}. \\
{\bf 7.} Comparison of the modeled light curves with the observation
implies that the mass of the white dwarf is about $1.2-1.3 M_\sun$,
the original mass transfer rate $(2-3) \times 10^{-6} M_\sun$~yr$^{-1}$,
the viscous timescale $20-40$ days, and the efficiency of 
stripping by winds $c_1= 5-10$. \\
{\bf 8.}  The on/off timescales of supersoft X-rays are as short as
one or two days \citep{rei00}.
Only the contraction/expansion of the white dwarf photosphere
cannot explain the rapid emergence/decay of supersoft X-rays 
because the contraction/expansion timescales of the white dwarf
photosphere are too long to reproduce such a rapid emergence/decay
of supersoft X-rays.
\par
     The white dwarf in \objectname{RX~J0513.9$-$6951} is now constantly
growing in mass at the critical rate of $\dot M_{\rm cr}$.
We expect that the mass of the white dwarf
increases up to near the Chandrasekhar mass and the white dwarf
eventually explodes as a Type Ia supernova.   Thus,
\objectname{RX~J0513.9$-$6951} is the first beautiful example
of the accretion wind
evolution, which is a key evolutionary process in a recently 
developed evolutionary path to Type Ia supernovae 
\citep[e.g.,][]{hkn96, hkn99, hknu99, hac01kb, lan00, lih97}.



\acknowledgments
     We are indebted to F. Meyer and Emi Meyer-Hofmeister
for stimulating us to work on \objectname{RX~J0513.9$-$6951}
during our stay at the Max-Planck Institute for Astrophysics.
This research has been supported in part by a Grant-in-Aid for
Scientific Research (11640226) from
the Japan Society for the Promotion of Science.

\begin{deluxetable}{lcrrrrll}
\tabletypesize{\scriptsize}
\tablecaption{Durations of high and low states
\label{high_low_states}}
\tablewidth{0pt}
\tablehead{
\colhead{$M_{\rm WD}$} & 
\colhead{$\dot M_{\rm MS}$} & 
\colhead{$t_{\rm vis}$} & 
\colhead{$c_1$} & 
\colhead{high} &
\colhead{low} &
\colhead{light} &
\colhead{comments} \\
\colhead{$(M_\sun)$} &
\colhead{$(10^{-6} M_\sun$~yr$^{-1})$} &
\colhead{(day)} &
\colhead{} &
\colhead{(days)} &
\colhead{(days)} &
\colhead{curve} &
\colhead{}
} 
\startdata
1.35 & 1.0 & 15.0 & 10.0 & 30 & 30 &  Fig.\ref{vmagfit_rxj0513_self_wdm1350}a & \\
1.35 & 2.0 & 15.0 & 10.0 & 40 & 40 &  Fig.\ref{vmagfit_rxj0513_self_wdm1350}b & \\
1.35 & 5.0 & 15.0 & 10.0 & 70 & 40 & Fig.\ref{vmagfit_rxj0513_self_wdm1350}c & fix $t_{\rm vis}$ \\
1.35 & 10. & 15.0 & 10.0 & 90 & 35 & Fig.\ref{vmagfit_rxj0513_self_wdm1350}d & \\
1.35 & 5.0 & 20.5 & 10.0 & 80 & 70 &  & \\
1.3 & 0.6 & 20.5 & 10.0 & 0 & $\infty$ &  & no winds \\
1.3 & 0.7 & 20.5 & 10.0 & 40 & 45 &  & \\
1.3 & 0.8 & 20.5 & 10.0 & 40 & 45 &  & \\
1.3 & 0.9 & 20.5 & 10.0 & 40 & 45 & Fig.\ref{vmagfit_rxj0513_self_acc}a & \\  
1.3 & 1.0 & 20.5 & 10.0 & 45 & 35 & Fig.\ref{vmagfit_rxj0513_self_acc}b & \\
1.3 & 2.0 & 20.5 & 10.0 & 65 & 40 & Fig.\ref{vmagfit_rxj0513_self_acc}c & \\
1.3 & 3.0 & 20.5 & 10.0 & 85 & 40 & Fig.\ref{vmagfit_rxj0513_self_acc}d & \\
1.3 & 5.0 & 20.5 & 10.0 & 110 & 40 & Fig.\ref{vmagfit_rxj0513_self}b &
 fix $t_{\rm vis}$ \\
1.3 & 7.0 & 20.5 & 10.0 & 130 & 30 & Fig.\ref{vmagfit_rxj0513_self_acc}e & \\
1.3 & 9.0 & 20.5 & 10.0 & 140 & 25 &  & \\
1.3 & 10. & 20.5 & 10.0 & 150 & 20 & Fig.\ref{vmagfit_rxj0513_self_acc}f & \\
1.3 & 15. & 20.5 & 10.0 & 160 & 2 & Fig.\ref{vmagfit_rxj0513_self_acc}g & \\
1.3 & 20. & 20.5 & 10.0 & $\infty$ & 0 & Fig.\ref{vmagfit_rxj0513_self_acc}h & modulation\tablenotemark{a} \\
1.3 & 5.0 & 14.0 & 10.0 & $\infty$ & 0 &  & modulation \\
1.3 & 5.0 & 16.0 & 10.0 & 95 & 15 &  & \\
1.3 & 5.0 & 18.0 & 10.0 & 100 & 25 &  & \\
1.3 & 5.0 & 23.0 & 10.0 & 115 & 50 &  & \\
1.3 & 0.7 & 36.0 & 5.0 & 75 & 50 & Fig.\ref{vmagfit_rxj0513_self_wdm1300_strip_c5}a & \\
1.3 & 1.0 & 36.0 & 5.0 & 85 & 40 & Fig.\ref{vmagfit_rxj0513_self_wdm1300_strip_c5}b  & \\
1.3 & 2.0 & 36.0 & 5.0 & 105 & 40 & Fig.\ref{vmagfit_rxj0513_self_wdm1300_strip_c5}c  & fix $t_{\rm vis}$ \\
1.3 & 3.0 & 36.0 & 5.0 & 120 & 30 & Fig.\ref{vmagfit_rxj0513_self_wdm1300_strip_c5}e & \\
1.3 & 5.0 & 20.5 & 5.0 & $\infty$ & 0 &  & modulation \\
1.3 & 5.0 & 35.0 & 5.0 & $\infty$ & 0 &  & modulation \\
1.3 & 5.0 & 45.0 & 5.0 & 165 & 40 &  & \\
1.3 & 5.0 & 50.0 & 5.0 & 175 & 60 &  & \\
1.3 & 2.0 & 45.0 & 5.0 & 115 & 65 &  & \\
1.3 & 2.0 & 40.0 & 5.0 & 110 & 50 &  & \\
1.3 & 2.0 & 34.0 & 5.0 & 100 & 35 &  & \\
1.3 & 2.0 & 33.0 & 5.0 & 95 & 30 &  & \\
1.3 & 2.0 & 30.0 & 5.0 & 90 & 25 &  & \\
1.3 & 1.0 & 62.0 & 1.5 & $\infty$ & 0 & Fig.\ref{vmagfit_rxj0513_self_wdm1300_strip_c015}a & no modulation\tablenotemark{b} \\
1.3 & 2.0 & 62.0 & 1.5 & 110 & 40 & Fig.\ref{vmagfit_rxj0513_self_wdm1300_strip_c015}c & fix $t_{\rm vis}$ \\
1.3 & 3.0 & 62.0 & 1.5 & 115 & 35 & Fig.\ref{vmagfit_rxj0513_self_wdm1300_strip_c015}d & \\
1.3 & 5.0 & 62.0 & 1.5 & 195 & 25 & Fig.\ref{vmagfit_rxj0513_self_wdm1300_strip_c015}e & \\
1.3 & 1.0 & 36.0 & 1.5 & $\infty$ & 0 &  & no modulation \\
1.3 & 2.0 & 36.0 & 1.5 & $\infty$ & 0 &  & modulation \\
1.3 & 5.0 & 36.0 & 1.5 & $\infty$ & 0 &  & modulation \\
1.3 & 2.0 & 60.0 & 1.5 & 150 & 35 &  & \\
1.3 & 5.0 & 60.0 & 1.5 & 190 & 25 &  & \\
1.2 & 1.0 & 32.0 & 10.0 & 80 & 45 & Fig.\ref{vmagfit_rxj0513_self_wdm1200}a & \\
1.2 & 2.0 & 32.0 & 10.0 & 120 & 45 & Fig.\ref{vmagfit_rxj0513_self_wdm1200}b & \\
1.2 & 3.0 & 32.0 & 10.0 & 155 & 40 & Fig.\ref{vmagfit_rxj0513_self_wdm1200}d & fix $t_{\rm vis}$ \\
1.2 & 5.0 & 32.0 & 10.0 & 200 & 30 & Fig.\ref{vmagfit_rxj0513_self_wdm1200}e & \\
1.2 & 3.0 & 35.0 & 10.0 & 160 & 50 &  & \\
1.2 & 5.0 & 35.0 & 10.0 & 210 & 40 &  & \\
1.2 & 3.0 & 33.5 & 10.0 & 160 & 50 &  &  \\
1.2 & 3.0 & 35.0 & 10.0 & 160 & 50 &  &  \\
1.2 & 5.0 & 35.0 & 10.0 & 210 & 40 &  &  \\
1.2 & 5.0 & 20.5 & 10.0 & $\infty$ & 0 &  & modulation \\
1.2 & 5.0 & 30.0 & 10.0 & 195 & 15 &  & \\
1.2 & 5.0 & 40.0 & 10.0 & 225 & 70 &  & \\
1.1 & 0.6 & 45.0 & 10.0 & 115 & 55 & Fig.\ref{vmagfit_rxj0513_self_wdm1100}a &  \\
1.1 & 1.0 & 45.0 & 10.0 & 120 & 50 & Fig.\ref{vmagfit_rxj0513_self_wdm1100}b &  \\
1.1 & 2.0 & 45.0 & 10.0 & 195 & 40 & Fig.\ref{vmagfit_rxj0513_self_wdm1100}d & fix $t_{\rm vis}$ \\
1.1 & 3.0 & 45.0 & 10.0 & 250 & 20 & Fig.\ref{vmagfit_rxj0513_self_wdm1100}e &  \\
1.1 & 2.0 & 40.0 & 10.0 & 175 & 25 &  & \\
1.1 & 2.0 & 50.0 & 10.0 & 205 & 65 &  & \\
1.1 & 5.0 & 20.5 & 10.0 & $\infty$ & 0 &  & modulation \\
1.1 & 5.0 & 40.0 & 10.0 & $\infty$ & 0 &  & modulation \\
1.1 & 5.0 & 60.0 & 10.0 & 380 & 70 &  & 
\enddata
\tablenotetext{a}{mass accretion rate $\dot M_{\rm acc}$ 
is modulated by winds}
\tablenotetext{b}{mass accretion rate is not modulated by winds but
constant in time}
\end{deluxetable}






\clearpage

\clearpage

\clearpage

\clearpage

\clearpage

\clearpage

\clearpage

\clearpage

\clearpage

\clearpage

\clearpage

\clearpage

\clearpage


\end{document}